 \documentclass[preprint,5p,times,twocolumn,authoryear]{elsarticle}
\usepackage{comment}
\usepackage{multirow}

\usepackage{amssymb}
\usepackage{amsmath}
\usepackage{lipsum}
\usepackage{longtable}
\usepackage{url}
\usepackage{tikz}
\usetikzlibrary{mindmap}

\journal{Astronomy $\&$ Computing}

\usepackage{xcolor}

\begin{document}

\begin{frontmatter}



\title{A review of Unsupervised Learning in Astronomy}


\author[sfo]{Sotiria Fotopoulou}
\affiliation[sfo]{organization={School of Physics,  HH Wills Physics Laboratory, University of Bristol},
            addressline={Tyndall Avenue}, 
            city={Bristol},
            postcode={BS8 1TL}, 
            state={},
            country={United Kingdom}}

\begin{abstract}
This review summarizes popular unsupervised learning methods, and gives an overview of their past, current, and future uses in astronomy. Unsupervised learning aims to organise the information content of a dataset, in such a way that knowledge can be extracted. Traditionally this has been achieved through dimensionality reduction techniques that aid the ranking of a dataset, for example through principal component analysis or by using auto-encoders, or simpler visualisation of a high dimensional space, for example through the use of a self organising map. Other desirable properties of unsupervised learning include the identification of clusters, i.e. groups of similar objects, which has traditionally been achieved by the k-means algorithm and more recently through density-based clustering such as HDBSCAN. More recently, complex frameworks have emerged, that chain together dimensionality reduction and clustering methods. However, no dataset is fully unknown. Thus, nowadays a lot of research has been directed towards self-supervised and semi-supervised methods that stand to gain from both supervised and unsupervised learning. 
\end{abstract}







\end{frontmatter}




\section{Introduction}
\label{sec:introduction}

\subsection{What is learning?}

Learning from astrophysical data consists of extracting knowledge by constructing a mapping between the high-dimensional space of the observables to a lower dimensional space through, either an inference tool, e.g. empirical photometric redshift estimation, or an assumed model, e.g. estimating physical parameters from spectra such stellar mass or metallicity. An additional goal, approached as a byproduct of the learning process, is the identification of outliers. These are sources that do not conform to the inferred mapping as measured by some appropriately defined distance or similarity measure. Outlier or anomaly detection is becoming more and more prevalent in exploiting large astronomical datasets in pursuit of transient sources (supernova, gravitational waves), rare sources, i.e, short-lived phases of otherwise normal sources, including planet detection \citep{Sarkar2022MNRAS.510.6022S}, and in pursuit of the unknown unknowns \citep{Lochner2021A&C....3600481L}. 

The process of applying machine-learning (ML) methods on any kind of data typically encompasses the following steps 1) data gathering and calibration verification, e.g. zero points for photometric data, flux and wavelength calibration for spectroscopy 2) pre-processing, i.e. imputation of missing values, whitening, normalization 3) optionally dimensional reduction 4) hyperparameter tuning 5) performance validation. The first section of Table \ref{tab:ml_steps_reviews} provides references to relevant reviews, that are not astronomy specific. 

Largely defined by the presence of ground truth during the training process, supervised and unsupervised learning come with their distinct advantages and disadvantages. The field of supervised learning has enjoyed significant advancements in the past decade, starting by implications of deep learning, and ranging all the way to modern day transformer architecture. Supervised learning is the mapping between an input space and, a known, ground truth. It can be used to perform both classification and regression. However, it is impossible to extrapolate beyond the properties of the training set. Contrary, unsupervised learning encompasses methods that crystallise neighbourhood relationships in a high-dimensional parameter space, and methods that go a step further into projecting these relationships into lower-dimensional spaces that preserve - to a certain extent - local or global distances. Hence, the distinct advantage is the opportunity to discover new attributes, or new categories of objects. However, even though continuous maps or ranking of the data can be produced, the main goal of unsupervised learning is grouping the data into clusters.

Alternative approaches that fall in between the supervised/unsupervised division have also been developed aiming to by pass the limitations imposed by limited labelled training samples. These approaches include Self-supervised learning, Semi-supervised learning. Other methods aim to abstract information from the learned features with aim to push machine-learning towards artificial intelligence. These approaches can be categorised under the general terms of Transfer learning, and Representation learning. The second half of Table \ref{tab:ml_steps_reviews} provides reviews on each of these approaches without being domain specific. We shall return to these approaches in Section \S \ref{sec:modernML}.

Before we delve into the subject of unsupervised learning in astronomy, we point out that a summary of the ML methods discussed in this work, corresponding acronyms, and references to original papers or reviews on the subject are given in Table \ref{tab:ML_dim_red_method_papers} and Table \ref{tab:ML_clustering_method_papers}.

\begin{table}[]
    \centering
    \begin{tabular}{c|c} \hline\hline
    Approach & Review \\ \hline
  Pre-processing & \citet{MAHARANA2022-data-pre-processing-review}\\ 
  Dimensional Reduction & \citet{Jia2022-dimensionality-reduction-review}\\   
  Representation learning & \citet{Bengio-2013-representation-learning-review} \\ 
  Anomaly detection & \citet{Bansal-2022-anomaly-detection-review} \\ \hline
  Self-supervised &   \citet{Rani2023-self-supervised-review}\\
  Contrastive learning & \citet{Huertas-Company2023RASTI...2..441H} \\  
  Semi-supervised learning & \citet{vanEngelen2020-semi-supervised-review}\\
  Weakly-supervised learning & \citet{Zhou-2017-weak-supervision}\\
  Transfer learning &\citet{zhuang2020-transfer-learning-review}  \\ 
  Domain adaptation & \citet{Farahavi-2021-domain-adaptation-review}  \\
  \hline\hline
    \end{tabular}
    \caption{Reviews on the process of learning from data.}
    \label{tab:ml_steps_reviews}
\end{table}

\subsection{A brief historical perspective}

Data-driven discovery is part of the astronomers' DNA.  Even though the notion of `Big Data' is a loosely defined, and ever evolving term \citep[see][for a review]{Kitching2016-big-data}, the significance and challenges associated to acquiring and processing large amounts of data has long standing recognition in astronomy. In the following, we give a very brief historical context of advances in hardware, software, and data availability that have influenced astronomy since the turn of the millennium. 

\subsubsection{Early digital astronomy (pre-2000)}

Among the first digitised data have been scans of photographic plates, that already amounted to a few terabytes of data. These include digitised versions of Schmidt plates offered through the SuperCOSMOS Sky Survey\footnote{\url{http://ssa.roe.ac.uk//}} \citep[SSS,][]{Hambly2001MNRAS.326.1279H,Hambly2004ASPC..314..137H-supercosmos}, access to plates obtained from German observatories through the Archives of Photographic PLates for Astronomical USE\footnote{\url{https://www.plate-archive.org/cms/home/}} (APPLAUSE), while some of the plates obtained in the USA can be found through the project Digital Access to a Sky Century \@ Harvard\footnote{\url{https://dasch.cfa.harvard.edu/}} (DASCH, \citet{Grindlay2012IAUS..285...29G}) and the Maria Mitchell Observatory\footnote{\url{https://www.mariamitchell.org/astronomical-plates-collection}}.

Pre-2000, major all-sky digital surveys where already in place, including the Two Micron All Sky Survey \citep[2MASS,][]{Skrutskie2006AJ....131.1163S}, collecting a staggering 25TB of imaging between 1997-2001, and software packages such as IRAF had emerged for wider use in the community \citep{Tody1986SPIE..627..733T,Tody1993ASPC...52..173T}. However, the wider astronomical community had limited access to computing resources, and data were collected and transferred as physical copies on hard drives and compact discs. Despite the challenges, we already have some applications of unsupervised learning on spectra classification, stellar classification and so on \citep[e.g.][]{Storrie-Lombardi1994VA.....38..331S, Hernandez-Pajares1994MNRAS.268..444H, Connolly1995-first-PCA, Lahav1996MNRAS.283..207L, Faundez-Abans1996A&AS..116..395F, Naim1997ApJS..111..357N,Bailer-Jones1998MNRAS.298..361B,Galaz1998A&A...332..459G, Tagliaferri1999A&AS..137..391T, Lee1999ITGRS..37.2249L,VilelaMendes1999LNP...522..257V}.

\subsubsection{The multiwavelength era (2000-2010)}
In the years 2000 -- 2010, multiwavelength astronomy came into focus with extragalactic surveys such as the \textit{Sloan Digital Sky Survey} \citep[SDSS,][]{Gunn1998AJ....116.3040G, York2000AJ....120.1579Y}, \textit{Classifying Objects by Medium-Band Observations in 17 filters} \citep[COMBO-17,][]{Wolf2001A&A...365..660W,Wolf2001A&A...377..442W}, the \textit{Cosmic Evolution Survey} \citep[COSMOS,][]{Scoville2007ApJS..172....1S}, to name only a few. At the same time, affordable desktop computers, and the rise of Web 2.0, enabled fast communication, data transfer and collaborations such as national Virtual Observatories and the International Virtual Observatory Alliance\footnote{\url{https://www.ivoa.net/}}, established in 2002.

Publications of that time start to take advantage the massive datasets, and in addition to source classification they also focus on data-mining and knowledge discovery in databases \citep[e.g.][]{Andreon2000MNRAS.319..700A,Xui2001ChA&A..25..120X,Odewahn2002ApJ...568..539O,Rajaniemi2002ApJ...566..202R,Turmon2002ApJ...568..396T,Hakkila2003ApJ...582..320H,Eyer2005MNRAS.358...30E,Wagstaff2005IPNPR.163J...1W, Hojnacki2008StMet...5..350H,Rudick2009ApJ...699.1518R,Marzo2009JGRE..114.8001M,Sarro2009A&A...506..535S,Barra2009A&A...505..361B}. \citet{Ball2010IJMPD..19.1049B} provide a review of astronomical applications of that era.

\subsubsection{Computing goes mainstream (2010-2015)}
 The information technology revolution has shaped a new future. Astronomy collaborations grow larger, with SDSS paving the way. By this point in time, most - if not all - universities have access to significant computing resources, laptop computers have become ubiquitous, while collaborative and open-source software development has given rise to the first release of Astropy \citep{AstropyCollaboration2013A&A...558A..33A} and Scikit learn \citep{scikit-learn}. \citet{Ivezic2014sdmm.book.....I} publish their book on \textit{Statistics, Data Mining and Machine Learning in Astronomy}, with associated code examples, ready to run on astronomical data.

Ease of data access and robust codes, lead to experimentation of several methods.  Many works focus in data clustering, classification, and outlier discovery \citep{Andrae2010A&A...522A..21A,SanchezAlmeida2010ApJ...714..487S,Coppa2011A&A...535A..10C,Varon2011A&A...531A.156V,Geach2012MNRAS.419.2633G,Richards2012MNRAS.419.1121R,D'Abrusco2012ApJ...755...92D,Way2012PASP..124..274W,Shamir2012JComS...3..181S,Shamir2013A&C.....2...67S,SanchezAlmeida2013ApJ...763...50S,Graff2014MNRAS.441.1741G,Krone-Martins2014A&A...561A..57K,CarrascoKind2014MNRAS.438.3409C,Damodaran2014AdSpR..53.1720D}.

\subsubsection{ML Revolution (2015-2020)}
The next five years changed the machine-learning landscape forever. The public release of Tensorflow \citep{tensorflow2015-whitepaper}, and PyTorch (2016) provided an easy-to-use Python interface to lower-level C++ code, able to run on GPUs. The collaborative nature of software development and hands-on training is further cemented by the introduction of Project Jupyter (2015) and Cloud computing (e.g. Google Cloud, Amazon Web Services).

Deep learning applications have created a profound change in the world. Data are getting ever larger; multi-terrabyte archives exist across wavelengths (VHS, GAIA, DES, etc), with petabyte and exascale astronomy in the works. Access to resources for testing is not an issue as hardware keeps getting better, and even more affordable.  Inference at large scale starts to become challenging.

Astronomy stays on top the newest methods with applications on morphology and image segmentation, and many papers showcase the strengths and weaknesses of algorithms applied on classification, dimensionality reduction and time-domain astronomy keeping in mind the challenges of upcoming ever larger datasets \citep[e.g.,][]{Fraix-Burnet2015FrASS...2....3F,Kim2015MNRAS.453..507K,Schutter2015A&C....12...60S,Huijse2015arXiv150907823H,Kugler2015MNRAS.451.3385K,Koljonen2015MNRAS.447.2981K,Tramacere2016MNRAS.463.2939T,Armstrong2016MNRAS.456.2260A,Sasdelli2016MNRAS.461.2044S,Tammour2016MNRAS.459.1659T,Lawlor2016ApJ...833...26L,Rubin2016ApJ...828..111R,Kuntzer2016A&A...591A..54K,Zitlau2016MNRAS.460.3152Z,Davies2016MNRAS.456.2183D,Speagle2017MNRAS.469.1186S,Wetzel2017PhRvE..96b2140W,Baron2017MNRAS.465.4530B,DehghanFiroozabadi2017PASP..129g4502D,Meingast2017A&A...601A.137M,Benavente2017ApJ...845..147B,Selim2017ExA....43..131S,Frontera-Pons2017A&A...603A..60F,Mislis2018MNRAS.481.1624M,Cabrera-Vives2018AJ....156..284C,Reis2018MNRAS.476.2117R,Garcia-Dias2018A&A...612A..98G,Hocking2018MNRAS.473.1108H,Reis2021A&C....3400437R,Giles2019MNRAS.484..834G,Garcia-Dias2019A&A...629A..34G,Kounkel2019AJ....158..122K,Ralph2019PASP..131j8011R}.

\subsubsection{Meta algorithms (2020-)}
In later years, the focus has shifted towards 1) developing meta-algorithms and pipelines that either incorporate several algorithms to achieve tailored data analysis 2) exploration of the latest ML methods such as contrastive learning and generative models. We witness an ever increasing interest in the latent space: its robustness, its interpretation, and its limitations \citep{Logan2020A&A...633A.154L}. Domain adaptation methods promise to transfer knowledge across samples, while likelihood free inference / simulation based inference are increasing in popularity, particularly due to their speed and flexibility \citep[e.g.,][]{vonwietersheimkramsta2024kidssbi,chen2023learning}. However, as complexity increases, model interpretability is lost. At the same time, physics-aware models try to capture some of the sought-after meaning in the data \citep[e.g.,][]{Xu2023arXiv231005227X,Moschou_2023-PINN}.

\subsection{Machine Learning in Astronomy}

Astronomy has, and will, remain enamoured with the possibilities of ML applications as the field faces several of the `Big Data' definitions \citep[volume, velocity, variety, veracity, etc;][]{Kitching2016-big-data}. Along with the very extensive literature, a number of very detailed reviews exist on various applications of machine-learning and data-driven discovery briefly summarised below. 

\subsubsection{Previous ML Reviews}\label{sec:other_reviews}

\citet{Ball2010IJMPD..19.1049B} provide an overview of data-mining and knowledge extraction from databases, referring to any organised collection of data, including also FITS files. This review provides useful advice on pre-processing caveats relating to astronomy applications from a practitioner's perspective, for example source association and masking, and discuss individual ML methods.

\citet{Fraix-Burnet2015FrASS...2....3F} reviewed ML methods used for classification specifically for extragalactic astronomy. The authors discuss both supervised and unsupervised methods for classification, describing commonly used methods and their applications in extragalactic astronomy. \citet{Huijse2015arXiv150907823H} focused on time-domain astronomy, and in particular on the challenges introduced due to the scale of the anticipated LSST Survey. They showcase that a combination of unsupervised, supervised, as well as an active learning approach, that injects domain knowledge when necessary, is mostly likely needed. 

\citet{Baron2019arXiv190407248B} summarised in a practical and pedagogical manner commonly used unsupervised and supervised methods, including performance metrics and instructive toy examples. \citet{ElBouchefry2020kdbd.book..225E} provide a brief historical overview of the nascent period of ML, cover definitions of various types of learning, and provide application in astronomy and geoscience. In \citet{Doorenbos2021inas.book..197D} the authors discuss a detailed comparison of outlier detection methods, based on SDSS data. 

Finally, \citet{Smith2023RSOS...1021454S} give an excellent pedagogical review on the use of neural networks in astronomy, covering developments from early years of using multi-layer perceptron to modern applications of auto-encoders and generative models. More recently, \citet{Huertas-Company2023RASTI...2..441H} provide a review of contrastive learning, a new approach of self-supervised learning.

\subsubsection{About this review}

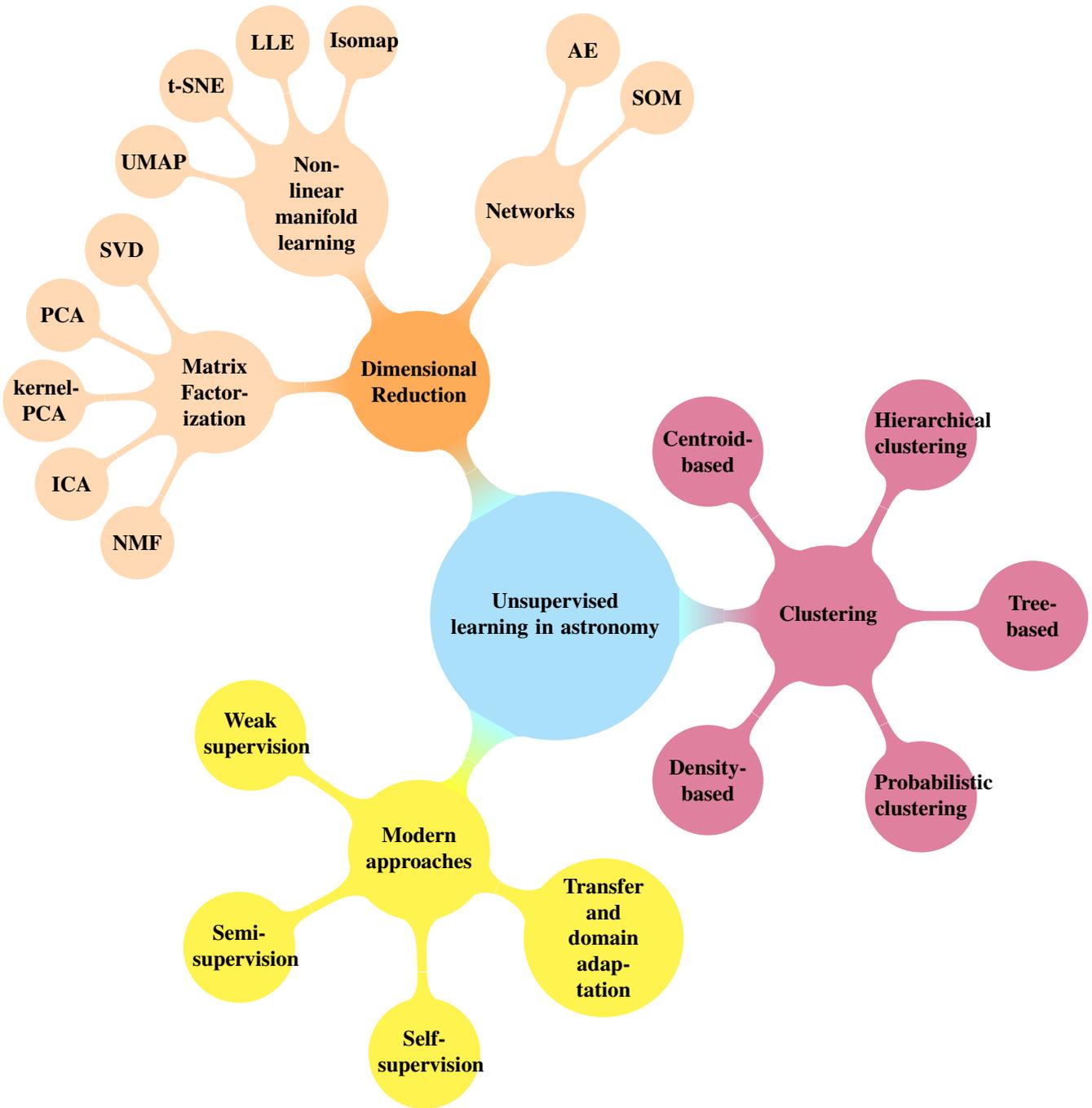
\begin{figure*}
\centering
\begin{tikzpicture}[mindmap, grow cyclic, every node/.style=concept, font=\sf\bf,
    concept color=cyan!30, scale=1.1,
    level 1/.append style={	font=\bfseries, level distance=4cm,sibling angle=120},
	level 2/.append style={	font=\bfseries, level distance=3cm,sibling angle=63},
 	level 3/.append style={font=\bfseries, level distance=2.5cm,sibling angle=30},]

\node{Unsupervised \\learning in astronomy}
child [concept color=yellow!80] { node {Modern approaches}
	child { node {Weak\\supervision}}
	child { node {Semi-supervision}}
	child { node {Self-supervision}}
	child { node {Transfer and domain adaptation}}
}
child [concept color=purple!50] { node {Clustering}
	child { node {Density-based}}
	child { node {Probabilistic\\clustering}}
	child { node {Tree-based}}
	child { node {Hierarchical\\clustering}}
	child { node {Centroid-based}}
}
child [concept color=orange!65] { node {Dimensional Reduction}
    child [concept color=orange!30] { node {Networks} 
    child {node {SOM}}
    child {node {AE}}
    }
    child [concept color=orange!30] { node {Non-linear manifold learning} 
    child {node {Isomap}}
    child {node {LLE}}
    child {node {t-SNE}}
    child {node {UMAP}}
    }
    child [concept color=orange!30] { node {Matrix Factorization} 
    	child  { node {SVD}}
    	child { node {PCA}}
    	child { node {kernel-PCA}}
    	child { node {ICA}}
    	child { node {NMF}} 
    }
};
\end{tikzpicture}
\caption{Graphical overview of unsupervised learning in astronomy.}\label{fig:graphical-review}
\end{figure*}

In this review, we will focus on the use of unsupervised methods in astronomy in the past 30 years. A broad search of astronomy and other sciences using the NASA/ADS database using the terms 'unsupervised learning', returned just over 1,000 items. We scanned through titles and abstracts for relevant literature and reduced the number to about 500 articles and book chapters. Some of the unrelated publications regarded student supervision and mentorship relationships, learning in terms of education, etc.

Of the 500 relevant papers, we tagged each entry by method used, and area of application. A number of the papers referred to Earth monitoring (including Moon and Mars) applications, volcanic activity, and modelling turbulent fluids. In the following, we will focus on the Astrophysical applications. Some of the Sun and solar system works will be mentioned briefly, but not discussed. We noticed a general trend of silos where communities used a specific code, or approach to model the corresponding data. Hopefully this review will spark some inspiration for testing other methods. The discussion presented in the remaining of this review is of course not limited to the retrieved $\sim$500 papers, which only served as a starting point.

A few major areas of application emerged during the review of the literature. In addition to the works on anomaly detection briefly mentioned earlier, the analysis of the GAIA survey is in the forefront of ML applications, including the use of unsupervised methods for stellar cluster detection. In extragalactic astronomy, heavy use of unsupervised methods is found in galaxy morphology in optical and radio images, source classification (e.g., star/galaxy/QSO), and spectroscopy.

In the remaining of this work, we will approach the subject neither in chronological order, nor in area of application or even by method. Rather we take the more abstract approach of examining the process of learning from astrophysical data. This includes first a discussion of traditional unsupervised approach including high-dimension space (\S \ref{sec:curse_dimensionality}) and features (\S \ref{sec:features}), dimensional reduction (\S \ref{sec:dim_redux}), data clustering (\S \ref{sec:clustering}). Finally, we briefly discuss very recent applications of self- and semi-supervised learning and domain adaptation (\S \ref{sec:modernML}).
We close with some recommendations on the way forward for future applications based on the observations made in the literature and our own experience (\S \ref{sec:recommendations}). 

We assume the reader is familiar with the nomenclature of machine-learning applications. For the readers that might need in depth explanations of the terminology, we point them to the excellent reviews mentioned previously in Section \S \ref{sec:other_reviews}.  Finally, we use the term `model' to refer to astrophysical models, as the spectral energy distribution of a galaxy and machine-learning models, as well as the algorithm and tuned hyperparameters created to describe a dataset, e.g. a trained self-organising map. The meaning of the term will be specified only when it is not clear from the context.

\section{Input Features and the Curse of Dimensionality}\label{sec:curse_dimensionality}

Astronomical features fall in three main categories: 1) observed 2) deduced based on a model or 3) deduced based on data-driven property of the dataset at hand. The first category includes for example fluxes and colours, spectra, and time series.  The second category includes properties found usually in a catalog produced with traditional methods, such a Sercic index, Gini index, metallicities, mass, etc. Finally, the data-driven derivation might include PCA components, features learned by a CNN, etc.   

Astronomical or other data live in a multi-dimensional space that is impossible to visualise directly if the parameter space spans more than three-five dimensions, e.g. by inclusion of colour and size gradients as extra dimensions. This parameter space is typically sparse and creates computational challenges. Even if the data could be on a narrow hyperplane, uncertainties will inevitably scatter the data above and below this hyperplane. The uncertainties are introduced not only by the instrumental limitations, but also by the intrinsic scatter of physical properties, which in addition might occupy a continuum and not distinct classes that might be found in everyday physical objects.  

Employing a clustering method directly on the high-dimensional space is rarely feasible or necessary, since we actually expect observed data to be correlated due to the underlying physical emission mechanisms. Thus, unsupervised learning usually starts by reducing the observed or deduced parameter space to a lower dimension space\footnote{Supervised methods can also benefit from dimensional reduction but will not be discussed here in detail.} (see Section \S \ref{sec:dim_redux}), usually into three to four dimensions \citep[e.g.][]{Sasdelli2016MNRAS.461.2044S,Logan2020A&A...633A.154L}, in which clustering analysis is performed (see Section \S \ref{sec:clustering}). The exact of number of dimensions is found through experimentation, by monitoring model performance \citep{Logan2020A&A...633A.154L} or iteratively through e.g. the average silhouette method as is the case for k-means.

\section{Pre-processing}\label{sec:features}

As is the case for supervised learning, unsupervised algorithms work best when the data have been cleaned and normalised before presented to the algorithm, e.g. by using the  \texttt{StandardScaler} in \texttt{Scikit.Learn}. This is common practice in the ML community. In the following, we note a few domain-specific issues relating to astronomical data.

\subsection{Data Imputation}
In ML, substituting missing data values with the mean of the distribution is standard recommendation. However, missing data in astronomy can arise due non-observed parts of the sky, due to technical artifacts (e.g. diffraction spikes), or due to non-detection in particular wavelengths and to a certain sensitivity (depth). The first two cases are non-informative, and can be usually made explicit in a catalog by the use of a placeholder value such as `-99'. The latter case of non-detections however, carries information on a per-source basis.

Astronomical catalogs do not always encode this missing information, or they might include a homogeneous image depth across the catalog, corresponding to the mean depth of the image. This approach is of course common practice due its simplicity. However, as data become larger, it is increasing impossible to resort to new source detection for each source of interest. Therefore, it would be beneficial for many analyses to include the measured flux at the location of the source across wavelengths.


\subsection{Normalization}
Many algorithms treat larger numerical values as more important, hence it paramount that this effect is removed before modeling the data. Standard ML practice includes the `whitening' and `normalisation' of the data, whereby the mean of the distribution is centered to zero and the standard deviation is scaled to one.

Importantly, this scaling has to be preserved and applied with the same scaled factor to new data that might be presented to a model at a future instance.

\section{Dimensional reduction}\label{sec:dim_redux}

Often, astrophysical features are correlated, and thus we can create a mapping between the sparse high-dimensional space and a lower-dimensional space that capture the majority of the information included in the data. A review of popular methods can be found \citet{Jia2022-dimensionality-reduction-review} and details on the methods can be found in \citet{Bishop2006}.

The new representation of the data in the lower-dimension space is called a \textit{latent space}. Most methods will attempt to create latent space compression that is capturing the most informative aspect of the data, which at the same time are usually non-intuitive to interpret. However, since they aim to reveal the information content on the observed data, they are often used as input in other downstream machine-learning methods, both in supervised and unsupervised tasks. In the following, we discuss popular dimensional reduction methods in astronomy, grouped by the algorithmic approach, i.e. matrix factorization (\S \ref{sec:matrix_factorisation}), manifold learning (\S \ref{sec:non_linear_manifold_learning}), and networks (\S \ref{sec:networks}).

\subsection{Matrix factorization}\label{sec:matrix_factorisation}

\subsubsection{Singular Value Decomposition (SVD)}\label{sec:svd}

Singular value decomposition (SVD) is the most general factorization of an $M\times N$ matrix. Namely, any matrix $\mathbf{A}$ can be re-written as:
\begin{equation}\label{eq:svd}
    \mathbf{A} = \mathbf{U}\cdot\mathbf{\Sigma}\cdot\mathbf{V^T}
\end{equation}
where $\mathbf{U}$ is an $M\times N$ column-orthogonal matrix, $\mathbf{\Sigma}$ is an $N\times N$ diagonal matrix, and $\mathbf{V}$ is another $N\times N$ column-orthogonal matrix. The diagonal elements of the $\mathbf{\Sigma}$ matrix are called the \textit{singular values} \citep[see][for a detailed discussion]{Press2007}.
SVD is used frequently for signal decomposition, with applications in astronomy ranging from decomposition of spectra of individual sources \citep[e.g.,][]{Simon1994A&A...281..286S,Piana1998A&AS..132..291P,Hobbs2006MNRAS.369..655H,Amara2012MNRAS.427..948A,Romano2017LRR....20....2R}, to cosmology and large scale structure \citep[e.g.,][]{Nicholson2010JCAP...01..016N,Vanderplas2009AJ....138.1365V, Bennett2013ApJS..208...20B,PlanckCollaboration2016A&A...594A..11P}. 
SVD is closely linked to the derivation of principal component analysis (PCA) discussed in the following section (\S \ref{sec:pca}), as it is computationally more attractive compared to solving for eigenvalues using the determinant of a matrix.

\subsubsection{Principal Component Analysis (PCA)}\label{sec:pca}

Principal component analysis  \citep[PCA,][]{Hotelling_1936_PCA} is the data transformation that identifies the directions of maximum variance in the data, i.e. the principal components are the eigenvectors of the covariance matrix of the data. It is therefore, a linear projection of N-dimensional data, into a d-dimensional space with $\mathrm{d<N}$. The majority of the information is usually captured in the first few components. The physical interpretation of the principal components requires further domain specific investigation.

Calculating the eigenvectors of the covariance matrix could be accomplished using the determinant of the matrix, but that would require the matrix to be square. A very educational description of the more computationally favourable application of SVD is given in \citet{Press2007, Ivezic2014sdmm.book.....I}. Briefly, the first step before searching for the eigenvectors is to remove the mean of the data, as would appear as the first principal component. Let our $M$ observations of $N$-features be described as an $M\times N$ matrix, $\mathbf{A}$. The covariance matrix is:
\begin{equation}\label{eq:covariance}
    C_A=A^T\cdot A
\end{equation}
Using the SVD factorization of eq. \ref{eq:svd}, we have:
\begin{equation}
    \begin{split}
    C_A & = A^T \cdot A \\
        & = (U\cdot \Sigma \cdot V^T)^T \cdot (U\cdot \Sigma \cdot V^T)\\
        & = V\cdot \Sigma \cdot U^T \cdot U \cdot \Sigma \cdot V^T \cdot\\
        & = V \cdot\Sigma^2 \cdot V^T
    \end{split}
\end{equation}
Therefore, the SVD of the covariance matrix leads to a quick and stable determination of the eigenvalues of matrix A, particularly in the regime of very large number of features. In practice, popular PCA algorithms, including the \texttt{sklearn} implementation use some version of the SVD approach\footnote{\url{https://scikit-learn.org/stable/modules/generated/sklearn.decomposition.PCA.html}}.

PCA has found an astonishing number of applications in astronomy. A NASA/ADS search at the end of 2023 returns 8.4 thousand referred papers. Likely the first application of PCA in astronomy has been the classification of stellar spectra by \citet{Deeming1964MNRAS.127..493D}. Further PCA uses in galactic applications include \citet{Storrie-Lombardi1994VA.....38..331S} where the authors showed that the first five PCA components of stellar spectra can lead to fast and accurate stellar type classification when used as input in a neural network. \citet{Deb2009A&A...507.1729D} performed a PCA decomposition of star lightcurves showing that PCA can provide a fast alternative classification of variable stars, as a first step before more detailed analysis. \citet{Krone-Martins2014-UPMASK} introduced the framework `Unsupervised photometric membership assignment in stellar clusters' (UPMASK) to identify stellar cluster members, applied by many authors on GAIA data. PCA is performed on the photometric data as the first step of the analysis, to identify stars with similar composition. \citet{Hayes2020MNRAS.494.4492H} used PCA to compress a model library with the aim to extract informative priors to speed up forward modelling of exoplanet spectra. \citet{Matchev2022PSJ.....3..205M} used PCA to explore the physical properties of the exoplanet spectra benchmark dataset of \citet{Marquez-Neila2018NatAs...2..719M}.

Examples of extragalactic astronomy applications include \citep{Connolly1995-first-PCA} who used PCA to produce a set of informative photometric filter combinations to infer photometric redshifts. \citet{Sulentic2000-PCA-eigenvector-1-qso} pieced together multiwavelength data to inform the interpretation of Eigenvector 1, the first principal component derived from QSO spectra as shown earlier in \citet{Boroson1992-PCA-Eigenvector-1-spectra}, linked to the Eddington ratio \citep{Marziani2003-eigenvector-1}. \citet{Wild2014-PCA-SED-classification} modeled \citet{Bruzual2003MNRAS.344.1000B} SEDs convolved with photometric filter curves and used PCA components, dubbed `super-colours' to classify the SEDs of galaxies into star-forming and passive. \citet{Lawlor2016ApJ...833...26L} used PCA and a number of other methods to map SDSS spectra to lower dimensions and explored the correlation of the projected space to physical properties of galaxies. More recently, \citet{Logan2020A&A...633A.154L} used PCA as means to reduce the input parameter space before performing star/galaxy/QSO classification.

\subsubsection{Kernel PCA}\label{sec:kernel-pca}
Kernel PCA \citep{Scholkopf_1998_kernel_PCA} was designed to overcome the limitations of linear projection of complex a data structures. In principle, mapping the data to higher dimensional parameter space would allow linear decomposition with PCA. However, the new space might need to be of too high dimensions, and the mapping function not easily known. The intractable computing difficulties are mitigated with appropriate choice of kernels which substitute dot products in feature space, with kernel functions. It is important to note, that, unlike PCA, kernel PCA does not allow the exact construction of the data but only an approximation.

This method has been applied as pre-processing step for detecting Type 1 SNe photometric classification \citep{Ishida2013MNRAS.430..509I} and lensed QSO \citep{Agnello2015MNRAS.448.1446A} using a gaussian radial base function as kernel. \citet{Xiang2017MNRAS.464.3657X} used kernel PCA with a gaussian radial basis function on LAMOST stellar spectra to estimate physical parameters of stars. \citet{Wang2020A&A...643L...9W,Amaya2020FrASS...7...66A,Irfan2021MNRAS.508.3551I} have applied kernel PCA on solar image observations, \citet{Papaefthymiou2022MNRAS.517.4162P} on mid-infrared spectra classification of Ultra Luminous Red Galaxies (ULRIGS) and QSO (gaussian kernel). 

\subsubsection{Independent Component Analysis (ICA)}\label{sec:ica}
Independent Component Analysis \citep[ICA - see][for a review]{HYVARINEN2000-independent-component-analysis-review} seeks to separate a signal into statistically independent components, e.g. overlapping sound signals, unlike PCA which aims to recover a representation that captures the most dominant components in the data. It is a linear transformation, applicable only on non-Gaussian data. If all components are gaussian, then the resulting mixture is also a symmetric gaussian distribution that does not contain enough information to disentangle the signals. A disadvantage of this method is that the number of components must be defined manually, which can be explored by inspecting the residuals. In addition, the importance of each component must be explored on the basis of domain knowledge and existing understanding of the system.

This method has been used to perform blind source separation, for example on mid-infrared Spitzer maps \citet{Meidt2014ApJ...788..144M} and Cosmic Microwave Background maps \citep{PlanckCollaboration2014A&A...571A...1P}.
\citet{Meidt2014ApJ...788..144M} used ICA to isolate the light of old stellar populations to derive accurate mass-to-light ratio conversions on the Spitzer Survey of Stellar Structure in Galaxies \citep[S4G,][]{Sheth2010PASP..122.1397S,PlanckCollaboration2016A&A...594A..11P-XI-SMICA}.
\citep{Cardoso2008-independent-component-analysis} extended the ICA approach to multi-spectral analysis of the Planck data \citep{PlanckCollaboration2014A&A...571A...1P}. 

ICA has been also used to separated signals in spectra. Indicative applications include the spectra of X-ray binaries \citep{Koljonen2015MNRAS.447.2981K}, exoplanets \citep{Waldmann2013ApJ...766....7W,Morello2019AJ....157..205M}, galaxies \citep{Lu2006AJ....131..790L,Allen2013MNRAS.430.3510A}, AGN \citep{Xu2007ApJ...670...60X,Richardson2014MNRAS.437.2376R}, QSO \citep{Temple2021MNRAS.508..737T}. \citet{Chattopadhyay2019PASP..131j8010C} used ICA as dimensional reduction method on 47 derived physical properties of galaxies as pre-processing before clustering, however they used PCA to find the number of needed dimensions.

\subsubsection{Non-negative matrix factorization (NMF)}\label{sec:nmf}
Non-negative matrix factorization \citep[NMF][]{Lee1999-NNMF} was developed as a method that is able to identify parts of objects, such part of a face. NMF mixing matrix is constrained to be strictly positive, contrary to PCA learns a holistic representation of the data and allows subtractions of the components to reconstruct an object.

The first application in astronomy was by \citet{Blanton2007AJ....133..734B} to derive k-corrections on galaxy spectra. They found that a basis of five components can be used to reconstruct the galaxy synthetic spectral library of \citet{Bruzual2003MNRAS.344.1000B}. Soon after, \citet{Allen2011MNRAS.410..860A} used NMF to create a basis of component that capture the information in QSO spectra. They created a new way to measure the `balcinity' index of a QSO by projecting broad absorption line QSO (BAL-QSO) spectra on the same basis. \citet{Koljonen2015MNRAS.447.2981K} found that NMF performed better over PCA and ICA on decomposing the spectrum of X-ray spectrum of the X-ray binary GX 339-4. They find that five components can provide a good description of the data, and linked the components to physical properties of the system.

\subsection{Non-linear manifold learning}\label{sec:non_linear_manifold_learning}

Manifold learning methods aim to capture neighbourhood relationships that exist in the input high dimensional space and create a mapping to a lower dimensional space that preserves these relationships, allowing to visualise of even perform clustering on the data. In this section, we discuss four such methods that are better suited for data exploration and visualisation. All methods below claim to tackle `short-circuit' effects, whereby a manifold with many folds, like e.g. the swiss-roll dataset, can lead to jumps from one fold to the other. See \citet{Meilua2024-manifold-stats-review} for a review on manifolds.

\subsubsection{Distance \& Divergence measures}\label{sec:distances}
The most commonly distance measures used in manifold learning are the Euclidean and geodesic distances.
The well known Euclidean distance gives the length of a line connecting two points:

\begin{equation}
    d=\sqrt{\sum(x-y)^2}  \;\;or\;\; d=||x-y||
\end{equation}

While the geodesic is the shortest distance between two points, which differs from the Euclidean distance on a non-flat space. 

Other useful measures include the Mahalanobis distance that gives the distance between a point and distribution, the Wasserstein distance that measures the `work' needed to transform on distribution to another, the Shannon divergence that measures the similarity between two probability distributions, the Kullback-Leibler (KL) divergence that measures the relative information content between two distributions, and the Procrustes alignment that is a comparison measure of shapes.

\begin{figure}
    \centering
    \begin{tabular}{c}
    \includegraphics[width=0.45\textwidth]{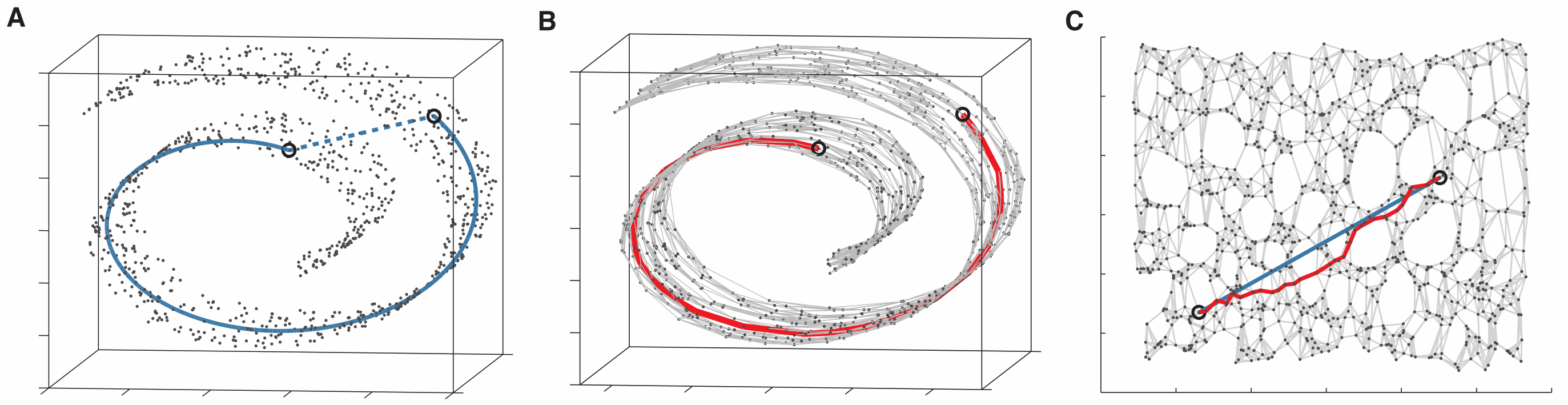} \\
    (a) \\
    \includegraphics[width=0.45\textwidth]{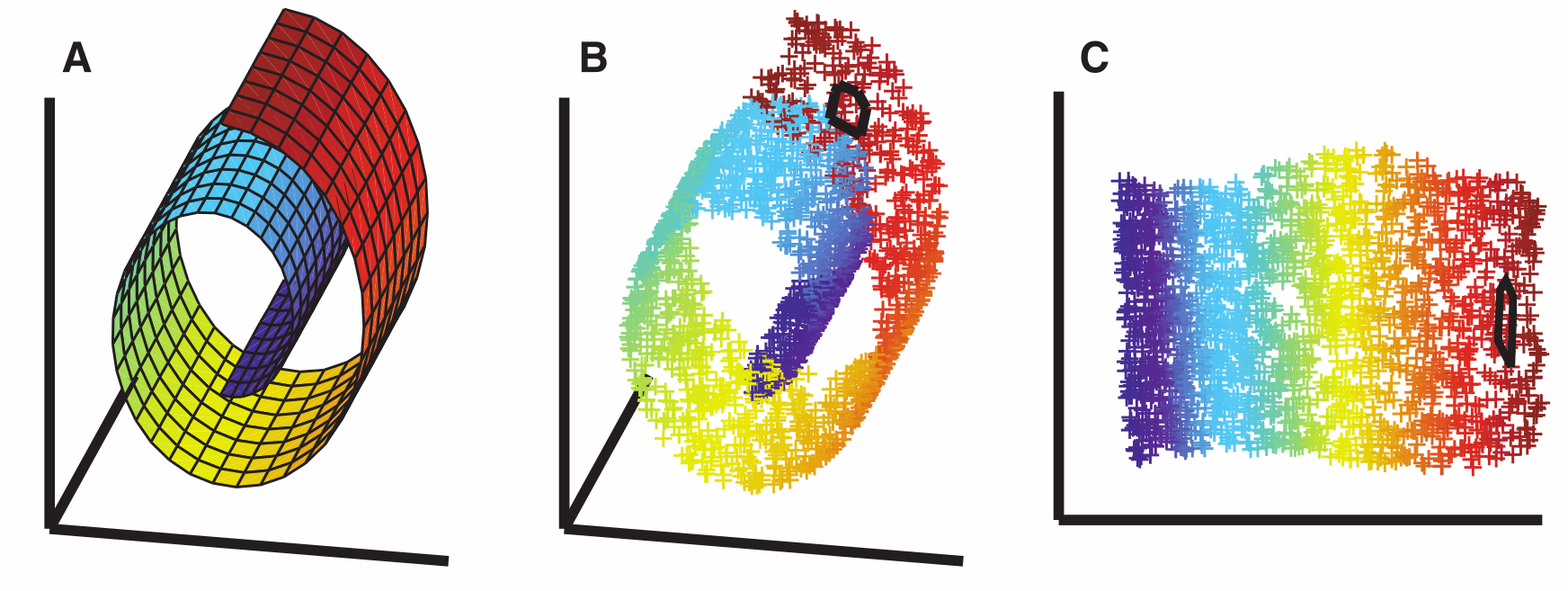} \\
    (b) 
    \end{tabular}
    \caption{Dimensional reduction with (a) Isomap, which finds the geodesic of neighbouring points, and b) Locally-linear Embedding, which find neighbourhoods where linear approximation holds. Both methods as showcased on the 'Swiss roll' dataset, where the curve of the manifold could cause a 'short-circuit' if a large neighbourhood is chosen. Fig. (a) adopted from \citet{Tenenbaum2000-ISOMAP}, fig. (b) adopted from \citet{Roweis2000-locally-linear-embedding}.}
    \label{fig:Isomap-LLE}
\end{figure}

\subsubsection{Isometric feature mapping (Isomap)}\label{sec:isomap}

Isomap \citep{Tenenbaum2000-ISOMAP} seeks to preserve global geometry of the input space, using geodesic distances. The algorithm operates in three steps 1) find neighbours within a certain distance (for computational efficiency) 2) create a graph and find minimum shortest path among all pairs 3) create a lower dimensional embedding for all points. The dimensionality of the data can be found by assessing the reconstruction error as a function of the number of projected dimensions. Generally speaking, Isomap can embed very complex manifolds, but can suffer from high computing time. A weakness of Isomap can be that the embedded points are too close to each other, and often organised in elongated clusters, making it non-trivial to apply clustering methods the data in the new space.

Isomap has found some applications in astronomy. \citet{Bu2014NewA...28...35B} used Isomap to classify SDSS DR9 \citep{Ahn2012ApJS..203...21A-SDSS-DR9} stellar spectra. They show that PCA and Isomap lead to different stellar subclasses. \citet{Sasdelli2016MNRAS.461.2044S} explored a wide variety of unsupervised learning methods to automatically classify Type Ia supernovae. They used Isomap to project to two dimensions the 4d latent space learned by an autoencoder. More recently, \citet{Matchev2022PSJ.....3..205M} explored dimensionality reduction techniques as means to find an embedding that will isolate the physical properties of exoplanet transmission spectra. Even though Isomap components capture physical properties such as molecule abundance and cloud opacity, they conclude that PCA is preferred as a more interpretable embedding.

\subsubsection{Locally Linear Embedding (LLE)}\label{sec:lle}

Locally Linear Embedding\footnote{Incidentally, Isomap and LLE appeared on the same journal side-by-side, with almost simultaneous submission and acceptance dates.} \citep[LLE,][]{Roweis2000-locally-linear-embedding} was designed to overcome the limitations of general linear projection methods (see Section \S \ref{sec:matrix_factorisation}), by creating local linear decomposition. The steps of the algorithm are largely similar to Isomap with the main difference on the second step. Namely, after finding the nearest neighbours of each point, instead of calculating the geodesic distances between the points as done in Isomap, LLE reconstructs the point in questions using a weighted sum of the neighbours. The fact that the algorithhm operates on small scales on the manifold, provides the flexibility to embed complex data. Contrary to PCA, LLE does not allow for the projection of new data once the manifold has been learned, however it can a powerful tool for data exploration, including the identification of outliers. \citet{Vanderplas2009AJ....138.1365V} presented the first application of LLE in astronomy, embedding the SDSS DR7 spectra \citep{Abazajian2009ApJS..182..543A} into two dimensions. They also presented a workaround for reusing the learned weight matrices of the neighbours in order to project new spectra. For the interested reader, \citet{Ghojogh2022arXiv221101369G} give a detailed overview of the LLE method and its many variants.

LLE applications in astronomy include embedding of stellar spectra \citep{Vanderplas2009AJ....138.1365V,Daniel2011AJ....142..203D,Bu2013PASJ...65...81B} and lightcurves of eclipsing binary stars \citep{Matijevic2012AJ....143..123M,Kirk2016AJ....151...68K,Bodi2021ApJS..255....1B}. \citet{Daniel2011AJ....142..203D} showed that the more than two dimensions are necessary to embed SDSS spectra, and in particular that most of the stellar spectra fall on a sequence in a 3d space (see their Figure 4).
\citet{Bu2013PASJ...65...81B} performed a comparison of LLE against PCA on M-type stellar spectra and found that LLE performs better, apart from the high noise regime.

\subsubsection{t-Distributed Stochastic Neighbor Embedding (tSNE)}\label{sec:tsne}

T-Distributed Stochastic Neighbor Embedding \citep[tSNE,][]{Maaten2008_tSNE} is another data visualisation method, developed to operate on non-linear manifolds. In particular, t-SNE models the likelihood of the data in the high dimensional space using a Gaussian distribution mapped to a Student-t distribution in the lower dimensional space. The long tails of the t-distribution solve the problem of over crowding  present in other methods. Due to the flexible representation of the data, t-SNE can learn more than one manifolds if they exist in the data. However, severe weaknesses of t-SNE include, 1) over-segmentation of the data in the lower dimensional space, i.e. creating more clusters than necessary 2) inability to project new data to an existing map 3) non-deterministic mapping 4) inability to map data to more than three dimensions, particularly problematic if the intrinsic dimension of the data is higher than that. Hence, the algorithm has been recommended since its inception to be used as data visualisation tool. However, parametric t-SNE \citep{vanderMaaten2009-parametric-t-SNE}, aims to tackle some of these issues.

Even though the weaknesses described above are clearly stated in the paper of \citet{Maaten2008_tSNE}, the astronomy community has made several attempts to use t-SNE as means for revealing clustering of data in the lower dimensional manifold. Often, the the projection is arbitrarily chosen to be two dimensional, which is not ideal for astronomical data. In addition, all t-SNE projections mentioned below that include a large number of objects suffer by increased complexity of the projected map making it impossible to group objects on the learned embedding. \citet{Kinson2021MNRAS.507.5106K} used probabilistic random forest to identify Young Stellar Objects (YSOs). In their Figure 13, they show clearly that the minority class they are interested in, cannot be blindly identified in the t-SNE projection.

\citet{Cotar2019MNRAS.483.3196C} used supervised and unsupervised methods to identify carbon-enhanced metal-poor (CEMP) candidate stars. The authors classified more than 600 thousand spectral from the GALctic Archaeology with HERMES pilot survey \citep[GALAH,][]{Duong2018MNRAS.476.5216D}. Their Figure 2, shows the clear limitations of trying to group similar objects together. Apart from a prominent group of CEMP stars identified manually, and with a supervised method, other CEMP stars are found spread over the t-SNE projection, and similarity to their neighbours is not evident. \citet{Steinhardt2020ApJ...891..136S} attempted to use t-SNE to identify quiescent galaxies. However, the map produced cannot be generalised. As the authors note, t-SNE needs to be recomputed every time new data need to be mapped. They chose to select a sample of galaxies with very narrow redshift ranges ($\mathrm{0.9<z<1.1}$, and $\mathrm{1.9<z<2.1}$). Their test sample sample is drawn from the same redshift distribution. However, if the map should be produced every time, and there exists a training sample, the approach is no different to k-nearest neighbours, since t-SNE finds objects with similar colours using a Euclidean distance. More recently, \citet{Youakim2023MNRAS.524.2630Y} used a known sample of $\omega$ Cen stars and a t-SNE projection of chemical abundances and stellar kinematics to identify members of the stellar stream. They defined a convex hull region on the t-SNE map based on known $\omega$ Cen stars and performed 100-fold bootstrap to identify the stars that fall within the pre-defined region, to combat the intrinsic stochasticity of t-SNE.

When few data are examined with a variety of classes present, t-SNE seems to perform well in grouping similar objects together. For example, \citet{George2017arXiv171107468G} created a 3d t-SNE map of the learned features of a convolutional neural network applied on the GravitySpy benchmark data. The modest size dataset (~8,500 elements) shows very well separated clusters. Contrary, t-SNE not only requires significant computing resources, but it also struggles to form distinct clusters as the number of points increases (roughly $>\;10^5$). In such cases, a t-SNE projection can be used as data mining tool, for example to search for outliers \citep{Giles2019MNRAS.484..834G, Webb2020MNRAS.498.3077W} or visualise the learned features of other methods \citep{Khan2019PhLB..795..248K,Chen2020A&A...643A.114C}. 

From astronomical works available in the literature using t-SNE, we can conclude a few general trends: 1) 3d projections seem to separate the data better than 2d projections 2) t-SNE cannot identify well minority classes 3) 2d maps can be used in conjunction with domain expertise to explore the map for anomalies  4) the shape of the embedded map is usually too complex for automatic clusterers (see section \S \ref{sec:clustering}). Often k-nearest neighbours can be used as an alternative if the aim is to search for similar objects based on a training set rather than searching for neighbours in the t-SNE projection.

\subsubsection{Uniform Manifold Approximation and Projection for Dimension Reduction (UMAP)}

\begin{figure*}
    \centering
    \includegraphics[width=\textwidth]{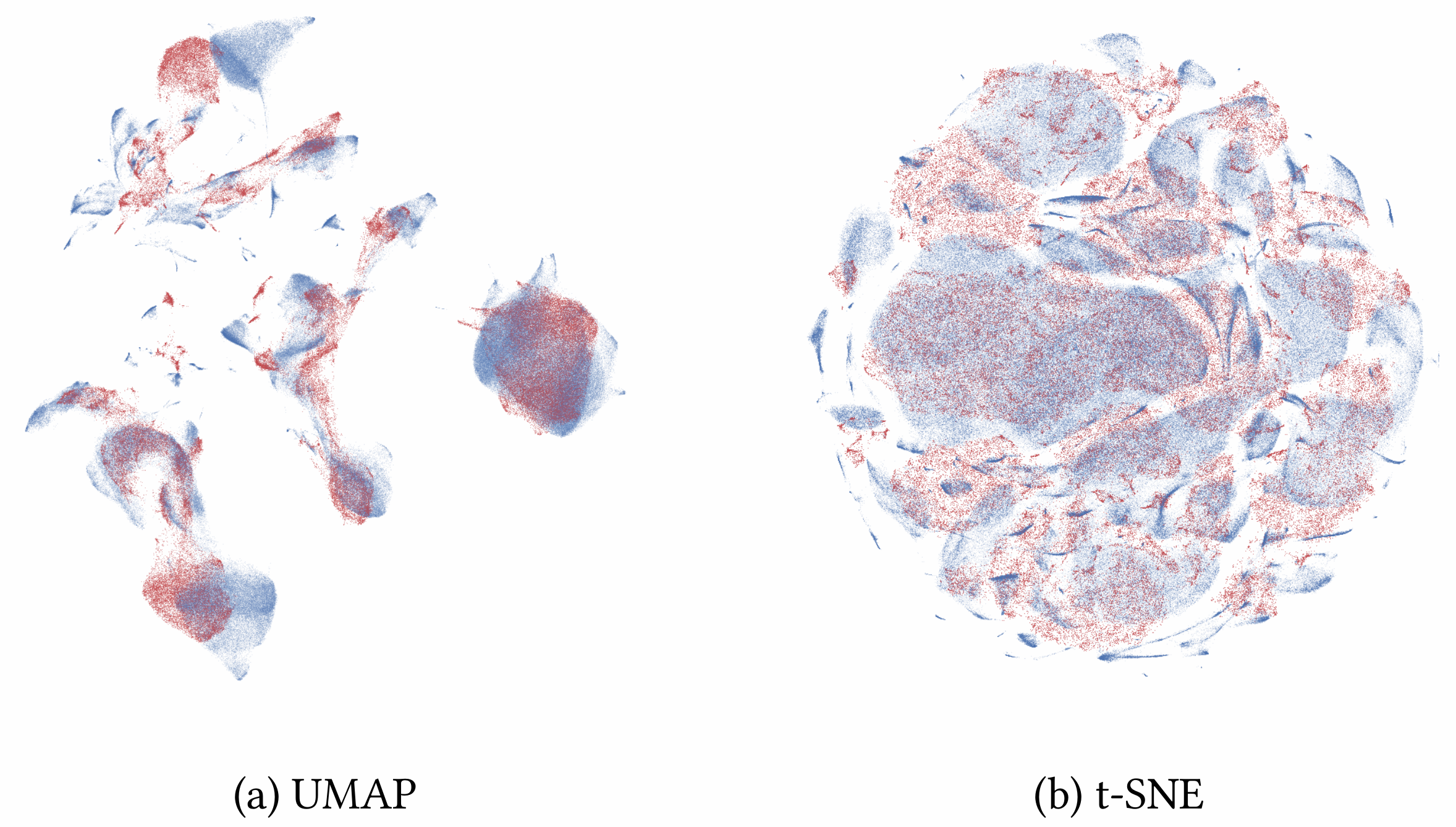}
    \caption{UMAP and t-SNE projections of 10\% subsample (red) and full (blue) flow cytometry dataset. If is visually clear that projection of new data on clustered spaced based on either of this projections leads to gross misalignment, here defined using Procrustes-based alignment. Figure adopted from \citet{McInnes2018_UMAP}.}
    \label{fig:UMAP-procrustes}
\end{figure*}

Uniform Manifold Approximation and Projection for Dimension Reduction \citep[UMAP,][]{McInnes2018arXiv180203426M-UMAP} is a more recent development in manifold learning. UMAP aims to better preserve the global structure of the data compared to t-SNE, with better scalability to larger datasets, and no restrictions on the number of dimensions. UMAP is based on the underlying assumptions that 1) the data are uniformly distributed on the manifold and 2) the underlying manifold is locally connected, i.e. there are no holes. As long as these conditions are satisfied locally, UMAP can be used to create a directed graph to model the k-nearest neighbours. 

Due to the stochastic approach on nearest neighbour search and gradient descent, UMAP, like t-SNE, does not create a deterministic mapping. This is a significant drawback when creating an embedding with subsample of the data in order to perform clustering in the lower dimensional space. Figure 7 of \citet{McInnes2018arXiv180203426M-UMAP} shows the alignment of various projected subsamples. Even though UMAP is more stable than t-SNE, a number of clusters are misaligned, hence it is not recommended to use any of the previous methods for classification, but rather only for data exploration and visualization. 

Astronomy works using UMAP span fields from molecules \citep{Lee2021ApJ...917L...6L}, to stars \citep{Sanders2023MNRAS.521.2745S} and galaxies \citep{Vega-Ferrero2023arXiv230207277V}, and across wavelengths.  \citet{Storey-Fisher2021MNRAS.508.2946S} searched for anomalies using a generative adversarial network. They used UMAP as a visualization tool, and they show that their detected anomalies tend to cluster together. \citet{Clarke2020A&A...639A..84C} used Random Forest to classify 111 million sources from SDSS. They showed that UMAP creates satisfactory clustering of spectroscopic data. However, we need to caution against assuming that the same trend generalises to photometric data (C. Logan private communication), as seemingly well separated classes gradually overlap and new, previously not seen clusters are created ad hoc. The same phenomenon is also demonstrated in Fig. 21 of \citet{Clarke2020A&A...639A..84C}.

\citet{Chen2022MNRAS.509.1227C} used UMAP to project 13 input features of about 600 fast radio bursts (FRB). UMAP creates nine, very distinct, clusters. This is a similar phenomenon that has been observed with t-SNE, namely the learned embedding can be informative when examining smaller datasets, while a large dataset ($>10^5$) will be mapped onto a continuous distribution, e.g. seen in the case of galaxies \citep{Clarke2020A&A...639A..84C,Storey-Fisher2021MNRAS.508.2946S, Slijepcevic2023arXiv230516127S}. This effect is seen in the UMAP representation of repeating FRB 20201124A in \citet{Chen2023MNRAS.521.5738C}. The authors expanded on their previous work, this including 1,745 FRBs. They find that the UMAP projection starts to become crowded. As has been noted with t-SNE, a 3d map perhaps would have been more informative. For example, \citet{Ricketts2023MNRAS.523.1946R} used a 3d UMAP projection, among other methods, to show the rich behaviour of lightcurve segments observed in the X-ray binary GRS1915.

\subsection{Networks}\label{sec:networks}

Even though technically networks fall under the non-linear manifold learning category, they deserve a dedicated discussion due to their different approach on modeling the manifold and their popularity.

\subsubsection{Self-organising maps (SOM)}\label{sec:som}

Self-organising maps \citep[SOM,][]{Kohonen1982_som}, or Kohonen networks, are an iterative dimensional reduction method based on competitive learning. To begin with, the geometry of the final map is chosen, e.g. an $\mathrm{N\times M}$ rectangle. Each cell on the map corresponds to a neuron with a randomly or otherwise (e.g. PCA) initialized weight vector. Each neuron is assigned to the most similar data point, hitherto known as the `best matching unit' (BMU). At the next step, the weight vector of the BMU and its neighbour are adjusted according to the assigned input data. The assignment between data and neurons is achieve through a distance metric, usually Euclidean.

To this day, SOMs continue to be very popular in astronomy and have found applications across all areas galactic and extragalactic astronomy, only a few can be highlighted here. Early attempts to demonstrate the applicability of SOMs on astronomical data include \citet{Hernandez-Pajares1994MNRAS.268..444H} which grouped the Hipparcos catalog of stars, along with synthetic data, to identify stellar components in the Milky Way (disc, halo), and \citet{Maehoenen1995ApJ...452L..77M-SOM} which a SOM to perform point-source detection, providing as simulated point-sources as input image stamps. 

Automated classification of stellar spectra has remained among the most researched topics, as it has traditionally relied heavily on human visualisation. SOM applications on stellar and galaxy spectra include \citet{Xue2001ChA&A..25..120X,Teimoorinia2022AJ....163...71T}, while lightcurve classification with SOM has been attempted by \citet{Armstrong2016MNRAS.456.2260A,Sasdelli2016MNRAS.461.2044S}. Another popular application of SOMs include morphological classification of optical \cite{Diaz-Garcia2019A&A...625A.146D,Holwerda2022MNRAS.513.1972H} and radio \citep{Galvin2020MNRAS.497.2730G,Mostert2021A&A...645A..89M,Gupta2022PASA...39...51G} images.

Photometric redshift estimation is the determination of galaxy redshifts based on they observed colours \citep[see][for a review]{Salvato2019NatAs...3..212S}. Redshift estimation with SOMs relied on the fact that galaxies with similar colours are expected to be at similar redshifts. Hence, a SOM map can be use to infer photometric redshifts of a galaxy populations with a restricted spectroscopic sample \citep{Geach2012MNRAS.419.2633G,Way2012PASP..124..274W,CarrascoKind2014MNRAS.438.3409C,Masters2015ApJ...813...53M,Speagle2017MNRAS.469.1186S,Suveges2017A&A...603A.117S,Wright2020A&A...637A.100W,Stolzner2023MNRAS.519.2438S}. Even though labels are needed to assign redshifts, this application is more akin to label propagation as the labels themselves are not part of the training. In a similar fashion, recently SOMs have been used to estimate physical parameters of galaxies \citep{Hemmati2019ApJ...881L..14H,Davidzon2022A&A...665A..34D}.

Finally, \citet{Rajaniemi2002ApJ...566..202R} used a SOM to investigate the three Gamma-Ray Burst (GRB) classes reported in \citet{Mukherjee1998ApJ...508..314M} (discussed in detail in Section \S \ref{sec:clustering}), and find no significant evidence for the existence of the third class \citep[see also,][for a discussion on sample incompleteness]{Hakkila2003ApJ...582..320H}.

\subsubsection{Auto-encoders (AE)}\label{sec:autoencoders}

\begin{figure}
    \centering
    \includegraphics[width=0.45\textwidth]{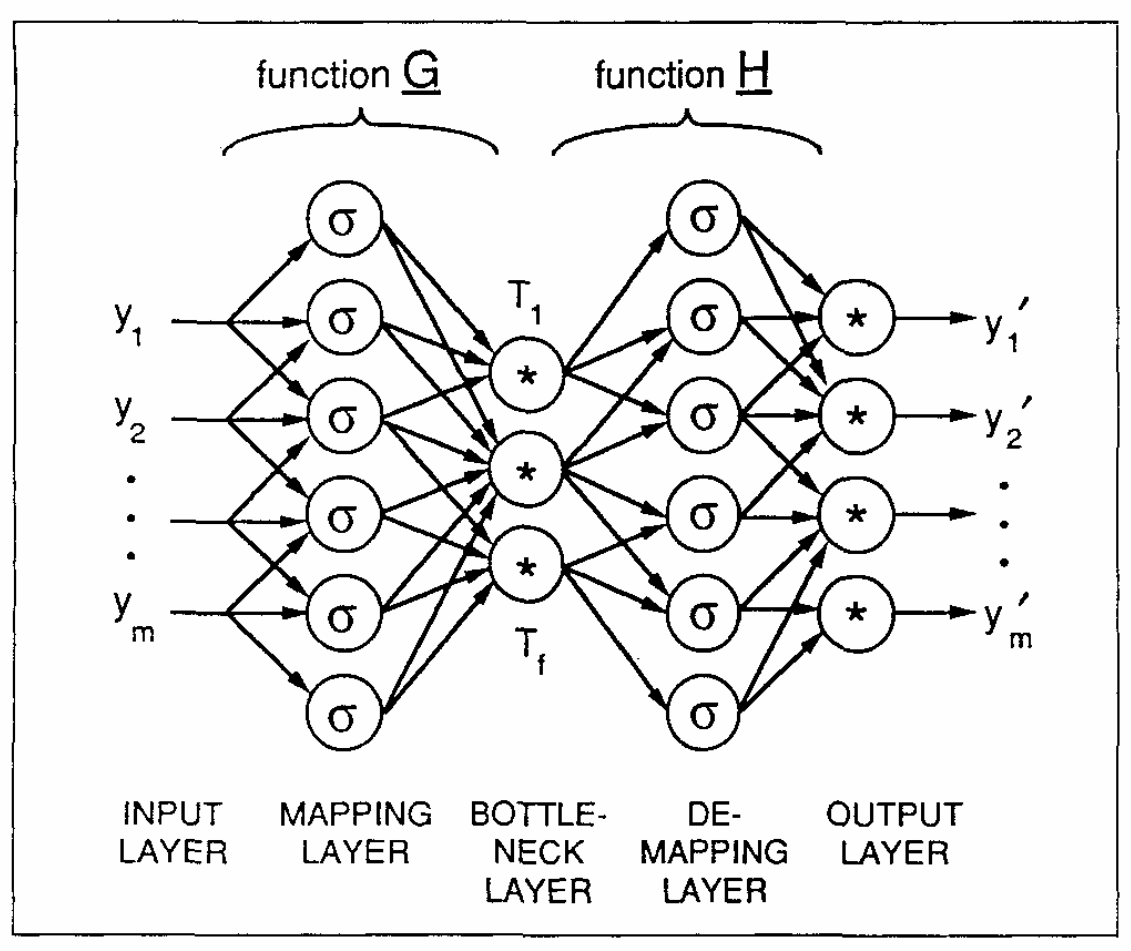}
    \caption{Autoencoder network, figure adopted from \citet{kramer1991_ae}.}
    \label{fig:autoencoder-krammer-1991}
\end{figure}

Neural-networks have been among the first ML methods to be embraced in astronomy \citep[e.g.,][]{Adorf1988LNP...310..315A,Odewahn1992AJ....103..318O,Lahav1995Sci...267..859L,Bertin1996A&AS..117..393B}. Auto-encoders \citep[AE,][]{kramer1991_ae,KRAMER1992_AE}, as their unsupervised counterpart, are receiving heightened attention as they are able to benefit from convolutions, useful for detecting features on images, while at the same time creating a reduced, informative, latent space. They comprise two components; first, the encoder part is a series of layers of progressively lower number of neurons, while the decoder starts from the narrowest part of the encoder, progressively expanding to layers of more neurons. The decoder part can be a mirrored architecture of the encoder. The narrowest part of the network is the latent space, also known as the bottleneck. AEs receive as input data which also form the target. Therefore, using the same training strategy of supervised neural networks, the weights and biases of the model are trained through backpropagation with the goal to replicate the input data minimising the reconstruction error. This means that the latent space carries enough information to be interpreted as a compressed view of the input.

Recently, AEs have received a lot of interest with further complex architectures being developed, including denoising Auto-encoders \citep[DAE,][]{Vincent2008-denoising-AEs} which learn to reconstruct a corrupted version of the data, convolutional Auto-encoders \citep[CAE,][]{Masci2011-convolutional-autoencoder} which use convolution layers in the encoder part of the AE, Variational Auto-encoders \citep[VAE,][]{Kingma2013-VAE} which substitute the bottleneck with gaussian distributions of which the mean and standard deviation are learned during the training. This means that the decoder can be used a probabilistic generative model.  On the other hand, Vector Quantised-Variatonal AutoEncoder \citep[VQ-VAEs,][]{van-den-oord-2017-vq-vae} substitute the latent random variables of VAEs with categorical variables, which have applications relating to speech and language.

The flexibility of AEs has inspired many astronomical applications. Some examples\footnote{See Section \S \ref{sec:bottleneck} for further examples regarding knowledge extraction.} include morphological classification \citep{Ma2019ApJS..240...34M,Chang2021MNRAS.503.1987C,Spindler2021MNRAS.502..985S,Tohill2023arXiv230617225T}, galaxy lens detection \citep{Cheng2020MNRAS.494.3750C}, physical parameter estimation \citep{Frontera-Pons2017A&A...603A..60F}, lightcurve classification \citep{Tsang2019ApJ...877L..14T}, outlier detection \citep{Liang2023AJ....166...75L,Han2022RAA....22h5006H}, among others.

\citet{Lahav1996MNRAS.283..207L} used 13 catalogued properties of the ESO-LV catalogue to identify galaxy classes. The input features included colours, parametric (de Vaucouleurs red and blue exponents) and non-parametric (light-ratios, asymmetry, surface brightness, etc) quantities. They show that the `encoder' network captures more information than linear projection with PCA. In the radio domain AE example applications include \citet{Ma2019ApJS..240...34M} classified radio AGN based on their morphologies and \citet{Mesarcik2020MNRAS.496.1517M} used AEs to reconstruct LOFAR data and assess the quality of the images. 

Further examples on physical parameter estimation include \citet{Frontera-Pons2017A&A...603A..60F} showed that the 2d latent space of a denoising AE applied on galaxies with redshifts $\mathrm{0.1<z<1}$ shows a clear distinction between the star-forming blue cloud and the passive red sequence of galaxies. \citet{Tsang2019ApJ...877L..14T} presented an architecture that extracts features, classifies, and performs anomaly detection on star lightcurves. 

AEs have also found application in the gravitational wave field. In particular, \citet{Sakai2022NatSR..12.9935S} use a VAE to extract features from the 2d time-frequency spectrograms provided by the Gravity Spy dataset \citep{BAHAADINI2018-gravity-spy} clustered later with other methods to identify groups of similar signals. On the other hand,  \citet{Moreno2022MLS&T...3b5001M} used AEs to learn noise data and identify true signals as outliers. Finally, \citet{Shen2017arXiv171109919S} used a DAE as part of larger ML framework to denoise gravitational wave signals, while \citet{Yang2023Senso..23.6030Y} used the Noise2Noise approach \citep{Lehtinen-2018-noise2noise} to suppress instrumental noise on gravitational wave data.

\section{Clustering}\label{sec:clustering}
Once the observations have been distilled into a space that carries the majority of in the information present in the data, the next step is the ranking of their similarity, leading to grouping of similar objects together. \citet{Jain1999-data-clutering-review} provide a review of core concepts and methods on ranking and clustering data. Their Figure 7 shows the taxonomy of the various methods commonly used. Detailed discussions on data clustering methods can be found in \citet{Aggarwal-data-clustering}.

In the following, we discuss some commonly used algorithms grouped by cluster identification philosophy: centroid based partitioning, hierarchical clustering, and density based clustering.

\subsection{Dissimilarities}\label{sec:dissimilarities}

We first define the notion of dissimilarity. Much like the distance measure was introduced for learning manifolds (see Section \S \ref{sec:dim_redux}), a pairwise dissimilarity measure is needed to assess if two points should belong to the same cluster. A common choice is none other than the squared distance between two points $x_i$, $x_j$ \citep{hastie_2009_elements-statistical-learning_book}:

\begin{equation}
    D(x_i, x_j) = \sum_{j=1}^{p} d_j(x_{ij}, x_{i'j}) = \sum_{j=1}^{p} (x_{ij} - x_{i'j})^2
\end{equation}

The extension of the pairwise dissimilarity to a group of points is introduced through the \textit{linkage}, which can be \textit{complete}, \textit{single}, \textit{average}, or \textit{centroid} \citep[e.g. see ][]{James-Intro-statistical-learning}. They each correspond to a property of the pairwise dissimilarities of two clusters; complete linkage is the maximum pairwise dissimilarity between two clusters, single linkage is the minimum dissimilarity, while average linkage is the average pairwise dissimilarity, and finally centroid linkage is the dissimilarity of the centroids.

\subsection{Centroid Based clustering}

Centroid or partition based algorithms (k-means and its variants) start by a predefined number of clusters to be found within the data. The data are assigned to the $k$ clusters through an iterative process which progressively minimizes the total sum of the distances between the data points and the center of their respective cluster. 

\subsubsection{k-means}

K-means \citep{macqueen1967_k-means, lloyd1982_k-means}, much like PCA, has been used in astronomy in numerous occasions. A heuristic partitioning of the data with this method is quick and easy to implement: we assign $n$ data points into $k$ clusters by minimizing the intra-cluster sum of squares with an iterative procedure. First, pick $k$ random locations that will act as the first guess of the cluster centers. Next, we calculate the Euclidean distance of each point from the cluster center, assign each point to its closest center. Recalculate the cluster means from the members. Repeat until the cluster center does not change any more. This algorithm works best when the clusters are well separated. The number of clusters can be deduced heuristically with the elbow method, i.e. by monitoring the sum of the distances as a function of number of clusters.

Due to the simplicity of the method, k-means has been applied to a variety of astrophysical applications, including stellar spectra \citep{SanchezAlmeida2013ApJ...763...50S,Garcia-Dias2018A&A...612A..98G,Garcia-Dias2019A&A...629A..34G}, the debate on the number of GRB classes \citep{Hakkila2003ApJ...582..320H}, clustering of X-ray spectra \citep{Hojnacki2008StMet...5..350H}, optical galaxy spectra \citep{SanchezAlmeida2010ApJ...714..487S} and so on.
In particular, \citet{SanchezAlmeida2010ApJ...714..487S} looked for $k$ clusters in $~900k$ SDSS spectra and highlight the importance of data normalisation. They note that if the data are not scaled to a common flux level, in their case the $g-band$, the classifier is driven by the flux of the source. They find that $k=17$ contains $99\%$ of the galaxies. In follow up work, \citet{SanchezAlmeida2013ApJ...763...50S} applied k-means clustering to the SEGUE star sample, obtained also with SDSS \citep{Yanny2009AJ....137.4377Y}, finding 16 classes in the data. 
\citet{Garcia-Dias2018A&A...612A..98G} applied k-means with $k=50$ on the APOGEE SDSS stellar spectra, and later merged manually the classes into nine groups, driven by measured physical properties (temperature, special gravity, and chemical abundance).

\subsubsection{k-medoids \& c-means}

Among the most rigid assumptions of k-means is the fact that the centre of the cluster can be an arbitrary point, and the fact that the membership assignment is strict. The k-medoids and fuzzy c-means algorithms are attempting to relax these conditions. These improvements have motivated applications in a wide variety of astronomical areas.

The k-medoids \citep{Kaufman1990-k-medoids} algorithm was an update to the k-means algorithm, that inspired many more optimisations \citep[e.g.][]{Ng2002-k-medoids-CLARANS, PARK2009-k-medoids}. Similarly to k-means, the expected number of clusters has to be determined by the user. However, contrary to k-means, the center of the cluster is one of the data points, and the distance of a data point to the center of the cluster can be any arbitrary similarity measure, not necessarily the Euclidean distance. Applications include globular cluster membership allocation \citep{Pasquato2019MNRAS.490.3392P} clustering of the latent space representations of galaxy morphologies to improve the Hubble sequence in a data-driven way \citep{Cheng2021MNRAS.503.4446C} and clustering of eclipsing binary light curves \citep{Modak2018arXiv180109406M}.

Fuzzy c-means \citep{BEZDEK1984-fuzzy_c-means} is a soft classifier. Contrary to k-means and k-medoids where a source can belong to only one cluster, c-means allows a \textit{fuzziness} which controls the degree up to which a source is permitted to belong to more than one clusters. This classification can be seen as a probabilistic assignment, a desirable trait for astronomical data. \citet{Barra2008AdSpR..42..917B} and \citet{Benvenuto2018ApJ...853...90B} used fuzzy c-means on  solar imaging and solar flare classification respectively, while \citet{Jamal2018A&A...611A..53J} applied the same method to assess the reliability of spectroscopic redshifts.

\subsection{Hierarchical clustering}

Hierarchical clustering methods do not use a predefined number of clusters. Instead, they use a dissimilarity measure and a stopping criterion to aggregate, or break up the data into clusters.

Agglomerative clustering \citep[AL][]{Johnson_1967_agglo_clustering} is a `bottom-up' clustering approach. Initially, each point is considered its own cluster. A dissimilarity measure is employed to merge clusters that are located close by, creating a dendrogram structure. There is an extensive list of linkage criteria that has been used in the literature. 
Divisive clustering is the inverse procedure compared to agglomerative clustering. It's a `top-down' approach, where the entire dataset is considered one large cluster, progressively split to sub-clusters based on a dissimilarity criterion. A dendrogram is created, starting by the objects that have the largest dissimilarity.

To name a few examples, hierarchical clustering has been applied on galaxy morphologies \citep{Hocking2018MNRAS.473.1108H,Martin2020MNRAS.491.1408M,Dai2023ApJS..268...34D}, as part of a hybrid neural network for point-source identification \citep{Andreon2000MNRAS.319..700A}, stellar type classification \citep{Garcia-Dias2019A&A...629A..34G}, and fast radio transients \citep{Aggarwal2021ApJ...914...53A}. \citet{Hojnacki2008StMet...5..350H} used agglomerative clustering as part of their clustering framework to identify the number of classes within the data, which were later retrieved with k-means.

\subsection{Minimum Spanning Tree (MST) or Friends of Friends (FoF)}

The Minimum Spanning Tree \citep[MST, see][for a historical review]{Graham-1985-minimum-spanning-tree-historical-review} is better known as the Friends-of-friends algorithm in astronomy, used to measure the clustering of galaxies in 2d and 3d space \citep{Press1982-FOF,Einasto1984-FOF}, which is outside the scope of this review. A MST is an acyclic graph that provides the shortest path between two points, hence it can be interpreted as the geodesic. Examples of MSTs applications relevant to source classification include refining the cluster membership fist found through k-means \citep{Cantat-Gaudin2019A&A...626A..17C}, and similarity search of supernova lightcurves \citep{deSouza2023A&C....4400715D}.
MST is used as part of Isomap (see Section \S), and (H)DBSCAN (see Section \S \ref{sec:DBSCAN}).

\subsection{Probabilistic Clustering}

Probabilistic clustering methods aim to model the observed data distribution as a random variable drawn from an underlying multivariate distribution \citep[][p. 61-86]{Aggarwal-data-clustering}. A very commonly used method is Gaussian Mixture Models \citep[GMM,][]{Dempster_1977_gaussian_mixture}. As the name implies, GMMs aim to approximate the underlying distribution that generated the observed sample as a mixture of $K$ gaussians. Thus, the clustering problem is transformed to an optimisation problem under which we are trying to find the optimal number of gaussians, and their means and variances that will generate best the observed data. A typical method to solve this optimisation problem is the iterative Expectation Maximization \citep[EM][]{Dempster_1977_gaussian_mixture, MCLACHLAN2015_Expectation_maximization_review} algorithm. The choice of the optimal model parameters can be done with information criteria, such as the Bayesian Information Criterion (BIC) or the Akaike Information Criterion (AIC), which penalize models with very high number of free parameters.

\citet{Meingast2017A&A...601A.137M} introduced \textsc{Pnicer}, a GMM to model the line of sight extinction of the interstellar medium.
GMMs have been used frequently to model the latent space of VAEs, for example to model X-ray Chandra data of Tycho's supernova \citep{Iwasaki2019MNRAS.488.4106I}, \citet{Cheng2020MNRAS.494.3750C} to model optical images with strong gravitational lenses, and \citet{Karmakar2018arXiv180901434K} to detect stellar clusters in images. We will revisit this type of application in Section \ref{sec:bottleneck}. 

GMMs have also been used to model galaxies, for example in terms of kinematics \citep[e.g.,][]{Ortega-Martinez2022MNRAS.516..197O,Du2019ApJ...884..129D,Du2020ApJ...895..139D} and as a population \citep{Fraser2023MNRAS.522.5758F}. In the latter, the authors modeled the IllustrisTNG-100 simulation as if was an observed population. They used GMM to extract three clusters of galaxies based on their broad band photometry.

\subsection{Density Based Clustering} \label{sec:DBSCAN}

Density-based Spatial Clustering of Applications with Noise \citep[DBSCAN][]{ester1996_dbscan}, and its hierarchical extension \citep[HDBSCAN][]{Campello_2013_HDBSCAN, Campello2015-HDBSCAN}, approach the presence of clusters within data from the perspective of minimum number of objects within a given radius. This definition can be applicable to any number of dimensions, but it does suffer from the curse of dimensionality which naturally makes the data sparse. However, since the input features can be often correlated, a workaround is to apply these methods following a first dimensional reduction of the feature space to about five dimensions \citep[e.g.,][]{Logan2020A&A...633A.154L}. 

\citet{ester1996_dbscan} used the concept of `core points' and `border points' to track the density of objects within a dataset. As the minimum number of neighbours must be predefined, we must take into account that border points will naturally have less neighbours. They define as `clusters' collections of points that are density-reachable within their respective clusters and as `noise' points that do not belong to any cluster. By tracking the connectivity of the points, DBSCAN is able to identify non-convex clusters, which is not the case for centroid algorithms. This is particularly of use in astronomy for spatial applications such as non-parametric galaxy morphology detection \citep{Tramacere2016MNRAS.463.2939T} and stellar streams \citep[e.g.,][]{Rudick2009ApJ...699.1518R,Kounkel2019AJ....158..122K}.

DBSCAN has been used in many areas of astronomy, ranging from galaxies \citep{Rudick2009ApJ...699.1518R,Tramacere2016MNRAS.463.2939T}, to planet detection \citep{Mislis2018MNRAS.481.1624M}, young stellar objects \citep{Prisinzano2022A&A...664A.175P}, pulsars \citep{Pang2018MNRAS.480.3302P}, and GRBs \citep{Abraham2021MNRAS.504.3084A}. A large amount of works have used DBSCAN to identify star clusters in physical space, with extensive application on GAIA data \citep[e.g.,][]{Castro-Ginard2018A&A...618A..59C,Garcia-Dias2019A&A...629A..34G,Castro-Ginard2019A&A...627A..35C,Noormohammadi2023MNRAS.523.3538N,He2023ApJS..264....8H,Alfonso2023A&A...677A.163A}.

HDBSCAN, as the hierarchical extension of DBSCAN, has found also a wide usage in recent applications. HDBSCAN uses minimum spanning trees which allow the discovery of clusters with varying density, contrary to DBSCAN that set the density as a constant across the entire dataset. The implementation of \citet{McInnes2017_HDBSCAN} provides also the detection of outliers in the data, and prediction support. 

Applications of HDBSCAN on star clusters and stellar streams dominate the astronomical literature \citep[see][ for a review of GAIA results]{Helmi2020ARA&A..58..205H}. \citet{Kounkel2019AJ....158..122K} used five dimensional input (galactic coordinates, parallax, and proper mottons) to search for stellar streams in the GAIA DR2 data, expanding previous work of \citet{Cantat-Gaudin2018A&A...618A..93C} to much larger scales. However, their find that the resulting recovered clusters depend on the algorithm configuration due to the fact that stellar clusters that are further away will naturally have a smaller extent on the sky, and smaller parallaxes. Their final catalog contains 1,901 individual clusters of about 288k stars, a very small fraction of the GAIA DR2 catalog. Further works searching for stellar clusters in GAIA DR3 have continue to provide a refined view of the Milky Way local neighbourhood \citep[e.g.][]{Moranta2022ApJ...939...94M,Gagne2023ApJ...945..119G}

 \citet{Logan2020A&A...633A.154L}, searched for optimal configuration to split the 100 million source KiDS dataset into stars, galaxies, and quasars. As mentioned earlier, they found that a dimensionality reduction to three to five dimensions was a necessary pre-processing step to be able to apply HDBSCAN successfully in photometric colour space. \citet{Webb2020MNRAS.498.5317W, Webb2021MNRAS.506.2089W} presented \textsc{ASTRONOMALY}, a framework for transient discovery and lightcurve classification. The main engine of their system is HDBSCAN clustering combined with Isolation Forest \citep[IF,][]{Liu2008_Isolation_Forest}. \citet{Aggarwal2021ApJ...914...53A} tested a number of unsupervised methods for classification of single pulse radio transients. They concluded that either DBSCAN, or HDBSCAN can be used for their application, with preference on DBSCAN. 

\section{Modern approaches to learning}
\label{sec:modernML}

Up to now, we have discussed commonly used approaches to unsupervised learning, namely dimensional reduction and clustering. Modern takes on learning, increasingly lean on model ensembles and frameworks that combine ML algorithms in sophisticated ways. The sharp distinction between supervised and unsupervised learning is becoming more blurred by approaches including self-supervised, semi-supervised, as well as transfer learning and domain adaptation. Relevant reviews on these topics are listed in Table \ref{tab:ml_steps_reviews}. In the following, we discuss some of the approaches found in the astronomy literature, nonetheless this is a very active field with interesting methods appearing regularly in the literature that could find applications to astronomy soon \citep[e.g.,][]{Berthelot2019arXiv191109785B-remixmatch,Yoon2020VIMEET}.

\subsection{Modelling the bottleneck}\label{sec:bottleneck}

\begin{figure*}
    \centering
    \includegraphics[width=\textwidth]{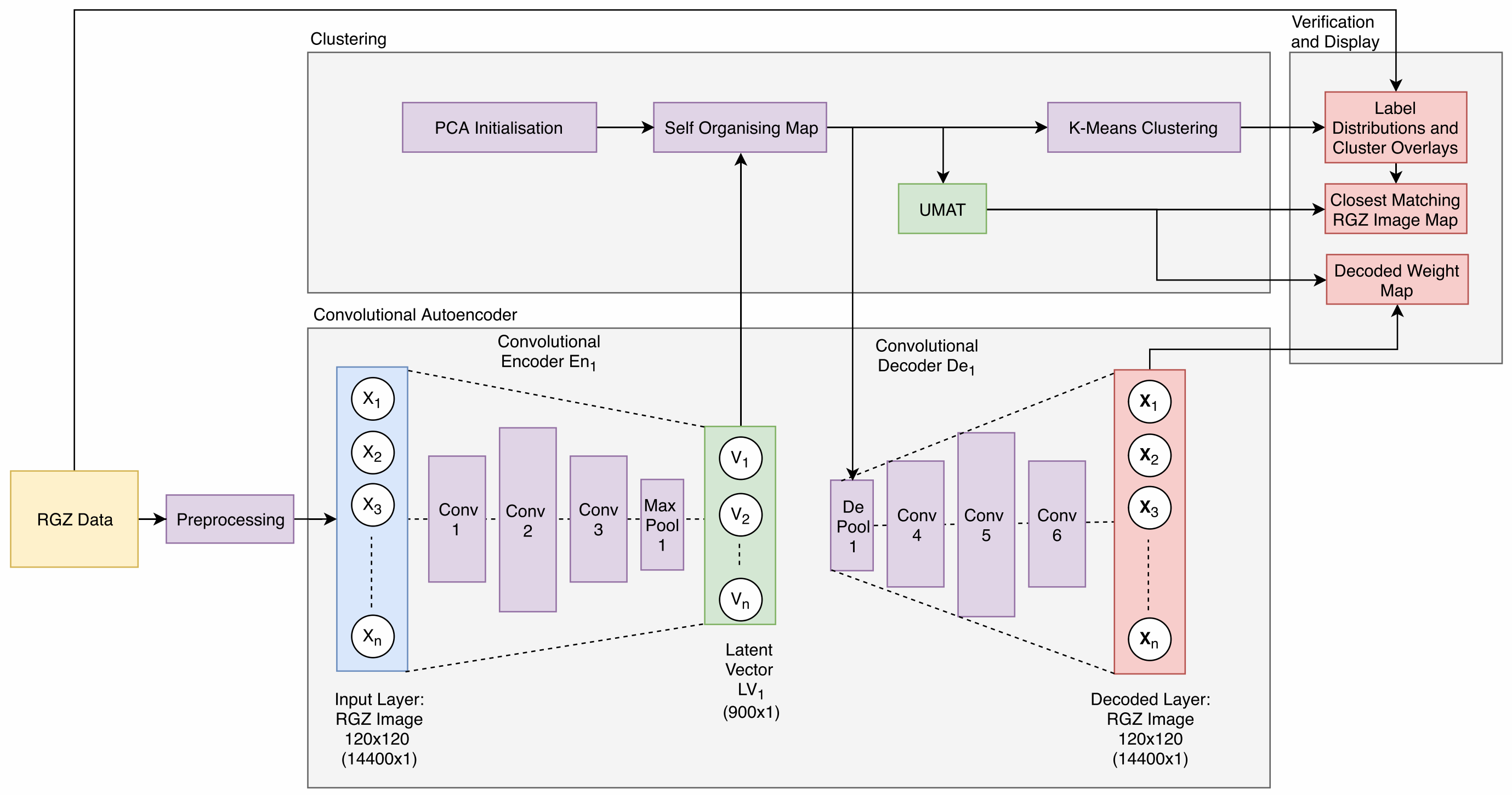}
    \caption{Complex frameworks are emerging in modern ML applications. Here we show one example of modelling the autoencoder latent space with a self-organising map, subsequently clustered with k-means. Figure adopted from \citet{Ralph2019PASP..131j8011R}.}
    \label{fig:AE-SOM}
\end{figure*}

Due to the compact representation of the data offered by AEs, several works have recently attempted to create a combination of architectures to model their bottleneck. These strategies take advantage of the non-linear projection of data, to the expense of feature interpretability. For example, \citet{Karmakar2018arXiv180901434K} modelled the bottleneck with Gaussian Mixture Models in order to identify stellar clusters in images. \citet{Villar2020ApJ...905...94V} used a combination of Gaussian processes for feature extraction, a recurrent-autoencoder for dimensionality reduction, and supervised random forest \citep{Breiman2001-random-forest} as their final classifier.
\citet{Tsang2019ApJ...877L..14T} used the latent space modelling of \citet{Zong2018-deep-GMM-UL-anomaly}, to classify variable star lightcurves.  The framework comprises a recurring neural network AE and a GMM to model the latent space. The GMM is further used to provide one-hot classification. Therefore, their dual-network approach classifies the data, and at the same time identifies anomalies. The authors find that their approach achieves accuracies comparable to supervised methods, for known classes.

\citet{Ralph2019PASP..131j8011R} (Fig. \ref{fig:AE-SOM}) used a series of algorithms to project radio images into progressively lower dimensions, ultimately clustered with k-means. However, as the authors discuss, significant assumptions are made when assigning hard boundaries of classes on a SOM, that should not be left without further investigation, especially keeping in mind that SOM cells do not form pure categories.

Recently, \citet{Forest2019-DESOM} proposed a method that combines an AE with a SOM at the bottleneck, training both at the same time. Inspired by this idea, \citet{Mong2023MNRAS.518..752M} developed a framework to classify `real vs bogus' transient sources. They find that the DESOM framework requires long training times, therefore they decoupled the AE and the SOM, by using a SOM projection which takes as input the latent space of the AE. However, the authors find that the current performance of their approach is not competitive against convolutional neural networks, trained on the same data. Even though the framework can be used as extra flagging, significant work is needed, in particular in extracting features from the images, and modelling the bottleneck.

\subsection{Weakly-supervised}
Weakly supervised is learning from noisy or poorly labelled data. In their review, \citet{Zhou-2017-weak-supervision} highlight three situations that lead to weak supervision. \textit{Incomplete} supervision reflects the fact that not all classes have assigned labels. This is a very common phenomenon in astronomy, as even the definition of boundaries between classes is non-trivial. \textit{Inexact} supervision corresponds to the fact that a label might represent a particular part of an images, e.g. morphology of the central galaxy, while neighbours are present in the same cutout. Finally, \textit{inaccurate} supervision refers to the fact that labels might be wrong. This is common occurrence in astronomy, whether the labels come from automated pipelines \citep[see ][for examples in SDSS and DESI spectroscopy]{Paris2018A&A...613A..51P, Alexander2023AJ....165..124A} or human annotators \citep[see][for examples of vote distributions among experts and volunteers alike]{Huertas-Company2015ApJS..221....8H, Walmsley2020MNRAS.491.1554W}.

All the above are present in astronomy data, with the added complication that the existing labels are more often than not a biased sample, e.g. mostly high signal-to-noise ratio. In the following, we discuss further only the methods related to unsupervised learning, relating to incomplete supervision.

\subsubsection{Semi-supervised}

Particularly true for astronomy, labelled training sets are bound to be not only the best quality examples but also typically well understood sources. Therefore, inferring on a larger sample leads to out-of-distribution problems. Semi-supervised learning blurs the line between the traditional unsupervised vs supervised split, by using both labelled and unlabelled data during training.

In their review, \citet{vanEngelen2020-semi-supervised-review} summarise the underlying assumptions that need to hold true for semi-supervised learning to work \citep[see also ][for detailed discussion]{Chapelle_2006_semi-supervised-review}. These are the following assumptions:
\begin{itemize}
    \item Smoothness: any two input points close to each other should lead to close by points in the target distribution.
    \item Cluster\footnote{An equivalent formulation is, the decision boundary should not cross a high density region.}: points belonging in the same cluster, should belong to the same class.
    \item Manifold: the data can be embedded in a lower dimensional manifold.
\end{itemize}
There is a number of strategies utilised for semi-supervised learning, largely split into \textit{inductive} and \textit{transductive} strategies. The former aim to create a predictive model, i.e. a model that can be applied to unseen data, such as a neural network. Contrary, the latter aims to answer the problem within the dataset available to the algorithm, e.g. find a lower dimensional embedding. The approach of combining known and unknown samples is standard practice in astronomy, therefore there is a number of applications of semi-supervised learning, including applications from AGN to supernova classification \citep[][to name a few]{Lawlor2016ApJ...833...26L, Villar2020ApJ...905...94V, Slijepcevic2022MNRAS.514.2599S}.

A related approach that is outside the scope of this review is the case of \textit{active learning}, which progressively incorporates the most uncertain prediction, as chosen by the model, drawn from the entire dataset. The data point is then presented to an \textit{oracle} and it is added into the training set, and the loop starts again \citep{settles-2009-active-learning,Stevens2021JOSS....6.3635S,Lochner2021A&C....3600481L,Walmsley2020MNRAS.491.1554W}.

\subsection{Self-supervised}

\begin{figure*}
    \centering
    \includegraphics[width=\textwidth]{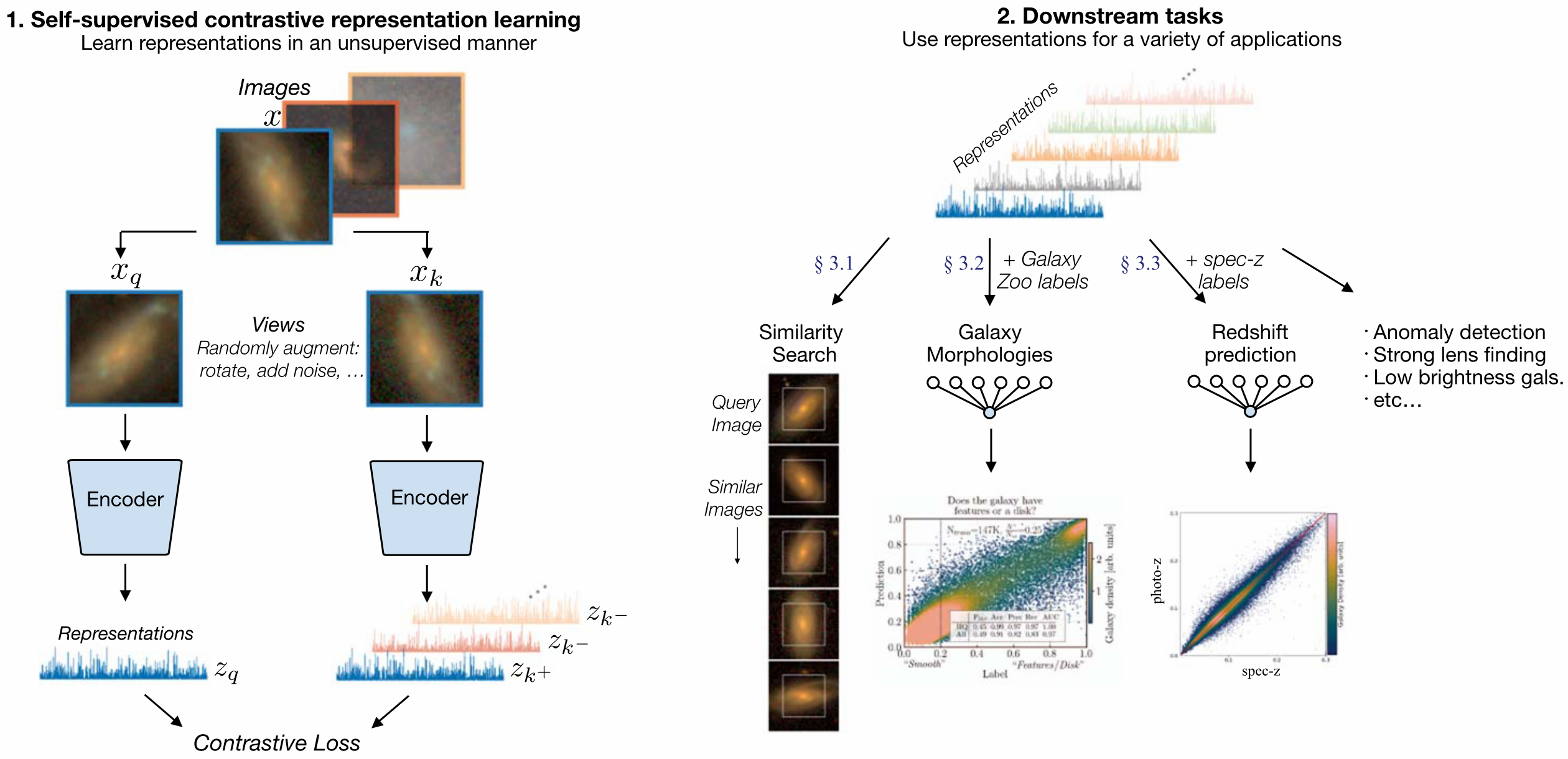}
    \caption{Application of contrastive self-supervised learning. Input features are augmented (rotation, flip, etc), and the contrastive loss is trained to bring representations of the same object closer in the latent space.  Figure adopted from \citet{Hayat2021ApJ...911L..33H}.}
    \label{fig:contrastive}
\end{figure*}

Self supervised learning learns first a pretext task, aimed to extract semantic information from the data. The pretext task comprises training a network to identify the original image based on transformed versions of the data (i.e. rotated, occulted, colourised, etc). The learned model is then used as initial guess when the downstream task is needed, i.e. classification or object detection. Two main approaches are used in the literature for pretext training: 1) contrastive learning, and 2) non-contrastive learning. In the first case, the algorithm is presented with variations of the positive example, such as zoom in and zoom out views of the image, cutouts, rotated, or scaled version of the original. At the same time some negative examples are also given to the algorithm. The pretext task, is to place the learned representations of the positive examples as close as possible in the learned latent space, while at the same time keeping the negative examples further away. This contrastive learning forces the algorithm to generalise the learned features of the positive class \citep[for a recent review see][]{Huertas-Company2023RASTI...2..441H}. The second non-contrastive approach uses only positive examples as input. This strategy involves `safety gates' such stop-gradient operations to ensure that the model will not suffer mode collapse, i.e. learn an embedding that is of lower dimension than required to describe the data.
Some of the works described previous (\S \ref{sec:bottleneck}), might fall under the non-contrastive learning, as long as the input data have been augmented. Since the learned representations form an abstract description of the data, they can be used input in further downstream tasks, which is a very desirable feature.

In the method \textit{A Simple Framework for Contrastive Learning of Visual Representations} \citep[simCLR,][]{Chen2020-simCLR} the authors highlight that the major reasons of the improved performance of their approach are due to 1) multiple data augmentations 2) non-linear transformation between the representation and the contrastive loss 3) normalized embeddings and 4) deep and wide architecture, large batch sizes, and longer training time. In a sense, the above observations offload the effort of labelling to computing time\footnote{The authors quote 1.5h of training, on 128 TPUs v3 cores for ResNet-50.}, which depending on the application might or might not be desirable.

\citet{Hayat2021ApJ...911L..33H} applied self-supervised contrastive learning on 1.2 million SDSS images (Fig. \ref{fig:contrastive}), by combining the architecture of \citet{Chen2020-simCLR}, and \citet{He_2020_CVPR}. They introduced augmentations relevant to astrophysical data, including galactic reddening and point-spread function smoothing. They used UMAP to visualise the 2048d representation, which shows that relative ordering of the learned representation corresponding to the galaxy morphology and orientation. They further showed that the learned representations can be used to assess galaxy morphology and estimate photometric redshift. They reported a training time of 12h, on eight NVIDIA V100 GPUs, for 50 epochs.

Since then, further works have explored the use of various contrastive-learning architectures, starting from galaxy images. To name a few, \citet{Stein2021arXiv211013151S} trained on a sample of 3.5 million galaxy cutouts from the Legacy Survey DR9 \footnote{\url{legacysurvey.org}}, using the momentum architecture of \citet{He_2020_CVPR}. Subsequently, they applied the trained network on 42 million galaxy from the same survey. The resulting representations were then used to identify similar objects. The dataset can be queried based on similarity on a public webpage\footnote{\url{https://github.com/georgestein/galaxy_search}}.
\citet{Sarmiento2021ApJ...921..177S} used spatially resolved images of 9,507 MaNGA galaxies and recovered the two-three major galaxy populations (passive, starforminng and intermediate) by performing k-means in the representation space. Other recent works that exploit contrastive learning on large galaxy samples include \citet{Wei2022PASP..134k4508W}, and \citet{Vega-Ferrero2023arXiv230207277V}.

On the other hand, \textit{Bootstrap Your Own Latent space} \citep[BYOL,][]{Grill2020_BYOL} is an example of the non-contrastive self-supervised approach. BYOL does not use negative examples during the learning phase, which would be used, e.g. in the case of simCLR, to avoid mode collapse. Instead the authors used two networks, the \textit{online} and \textit{target} neural networks which are built to interact and learn from each other. The online network, predicts the target network's representation of a different augmentation of the same image. The learning is refined by iterating the procedure flipping between the online and target networks. This method has been shown to outperform simCLR on everyday images, and it is slowly gaining popularity in astronomy \citep{Guo2022MNRAS.517.1837G, Slijepcevic2024RASTI...3...19S}.

\subsection{Transfer learning \& Domain adaptation}\label{sec:domain_adaptation}

Another approach to overcome the bottleneck of obtaining large labelled data and working with out-of-distribution data is transfer learning. Domain adaptation is a subset of transfer learning, whereby the objective is to match a \textit{source} distribution, $\mathcal{S}$, that contains labelled data, to a \textit{target} distribution, $\mathcal{T}$, that might be  data not drawn from identical distribution as the source. The goal of domain adaptation is to bridge the difference between the source and target distributions, often called a \textit{shift}. As the ultimate goal is to link the distribution $p(x,y)=p(y)p(x|y)$ in the source $p_{\mathcal{S}}(x,y)$ and target domains $p_{\mathcal{T}}(x,y)$, we can expect one of three discrepancies, prior shift, covariate shift, data drift, based on the origin of the disagreement. Table \ref{tab:domain_shifts} summarises the shift types, and how they relate to the marginal distribution $p(x)$, the prior, $p(y)$, and the conditional distribution $p(y|x)$. \citet{zhuang2020-transfer-learning-review} give a structured description of the various methods and strategies available in the ML literature, split mainly into instance-based and feature-based, while \citet[][]{Farahavi-2021-domain-adaptation-review} give an overview of domain adaptation approaches in shallow and deep learning.

\begin{table}[]
    \centering
    \begin{tabular}{|c|c|c|c|}\hline
        $p_{\mathcal{S}}(x)$ &  $p_{\mathcal{S}}(y)$ & $p_{\mathcal{S}}(y|x)$  & \multirow{3}{*}{Shift type}  \\
         vs & vs & vs &  \\
         $p_{\mathcal{T}}(x)$ &  $p_{\mathcal{T}}(y)$ & $p_{\mathcal{T}}(y|x)$ &  \\ \hline
          & $\neq$ & $=$ & Prior shift \\
          $\neq$ & & $=$ & Covariate shift \\
          $=$ & & $\neq$ & Data drift \\
          $\neq$ & $\neq$ & $\neq$ & intractable \\
          \hline
    \end{tabular}
    \caption{Possible reasons for domain adaption and necessary conditions for training.}
    \label{tab:domain_shifts}
\end{table}

Transformation equations are being used to, e.g. map instrument photometric systems to AB magnitudes, or to map from one standard photometric system to another \citep{Bessell2005ARA&A..43..293B}. Such transformations are a precursor of domain adaptation, which is nascent field in its modern form in astronomy. Recent applications include unsupervised domain adaptation of stellar spectra \citep{O'Briain2020arXiv200703112O}.
They used the \texttt{UNIT} framework \citep{Liu2017UnsupervisedIT} which creates a common latent space created with a VAE, trained under adversarial conditions. \citet{O'Briain2020arXiv200703112O} created two latent spaces, one for synthetic and real data and one only for real data. They found that the shared latent space was key in linking between data and synthetic spectra. Other recent applications include the prediction of physical parameters using simulation-based inference with a variety of methods, \citep[kernel PCA,][]{Gilda2021ApJ...916...43G}, including supervised approaches that are outside the scope of this review.

\citet{Ciprijanovic2023MLS&T...4b5013C} created a universal domain adaptation framework, \textit{DeepAstroUDA}, using semi-supervised learning, with application on galaxy morphology. In their paper, the give an excellent overview of domain adaptation methods relevant to astronomy. Their method combines the key ingredients of cross-entropy clustering to cluster source data, adaptive clustering to associate unlabelled data to labelled examples, and finally entropy loss to separate emerging new categories from data, including also outliers. They show that their method transfers knowledge successfully in the feature space, applicable in real-world and astronomical data.

We are confident that domain adaptation will find soon many more applications in astronomy, as we embrace \textit{simulation-based} or \textit{likelihood-free inference}.

\section{Recommendations}\label{sec:recommendations}

Below are a few general recommendations based on past experience, and many discussions with ML practitioners. See also Buchner et al., sub. for recommendations to newcomers in the field of ML in astronomy.

\subsection{Appropriateness} 
Machine-learning should be used only when necessary\footnote{The author has heard the question: `What shall I do when I have only a few data?'}. Certain ML approaches can be used as alternatives to traditional methods, such as least squares minimisation, keeping in mind that they will carry all the prior assumptions and biases of those methods. ML methods can be used also as exploration tools, such as finding the number of classes in the data, or looking for outliers. The pitfall in such exploration is that we always get back an answer. This means that we might look for distinct classes when a continuous distribution is more appropriate, or use a very biased sub-population to draw conclusions for the parent distribution. Manifold learning methods such as tSNE and UMAP are particularly prone to hallucinate new clusters.

\subsection{Benchmarks}
Computer science practitioners have invested many years to create benchmark datasets (MNIST, CIFAR, etc). Astronomy applications are far from such widely used benchmarks, probably for good reasons. However, data challenges have emerged for specific applications that pit codes against each other \citep[e.g.,][]{Hildebrandt2010A&A...523A..31H,EuclidCollaboration2020A&A...644A..31E,Savic2023ApJ...953..138S,Hartley2023MNRAS.523.1967H}. Such data challenges, while very useful, need to be interpreted with a pinch of salt when run on simulated datasets, due to domain shifts expected between data and simulations (see Section \ref{sec:domain_adaptation}).

\subsection{Visualisation}

\textit{Input space:} Topcat \citep{Taylor2005ASPC..347...29T} is an extremely powerful data exploration tool. We recommend to always plot projections and 3d plots of the input parameter space. Astronomy data are full problematic artifacts that creep into the final products, even for the most sophisticated pipeline. Data correlation matrices, when the number of dimensions allows, are very useful to drop extraneous features.

\textit{Latent space:} When examining a latent space, it is important to not over interpret structures, critically important when examining the latent representation of a labelled sample (e.g. spectroscopic sample of galaxies). As shown in Figure \ref{fig:UMAP-procrustes}, a latent space of a subsample might show different substructure compared to the parent sample.

\textit{Output:} Confusion matrices are very informative, especially when more than two classes are involved. It is good practice to write the number of objects per cell in addition to the colourbar. We recommend that the colourbar is scaled fixed to the range [0-1]. We motivate the field to adopt as standard practice to assign uncertainties in the model performance. Random seeds should be kept fixed during development, and mentioned in publications. However, they can induce significant variation in the trained model when left free.

\subsection{Generalisation}

New ideas must necessarily pass through a phase of experimentation and exploration before they are widely accepted. ML is no different. The enthusiasm of getting closer to `intelligent systems' and `knowledge discovery from data' is palpable. However, it is paramount that we distinguish between theory and practice. Many works have used simplified, simulated data to demonstrate zeroth order feasibility of a methods. The reality of astronomical data does not stop there. Missing data, either not observed or not detected, noisy data, instrumental effects, and so on, need to be incorporated as part of the modelling, either directly in the algorithm (e.g. bayesian neural networks) or through bootstraping. 

It is worth noting, that some areas such as gravitational waves are in the exact opposite situation compared to galaxy evolution. Namely, gravitational waves benefit from a strong theoretical foundation and are currently in the process of collecting larger observed samples. The approach and goal in this case is different, where ML is used as part of the data reduction (e.g. denoising) or signal detection as anomalies, and not for clustering.

\subsection{Barriers to discovery}

We recommend to bravely explore higher dimensional spaces and prioritise continuous distributions over classification whenever possible. It is tempting to refine the Hubble morphology sequence of galaxies using a data-driven approach \citep[e.g.][for an interesing discussion]{Cheng2021MNRAS.503.4446C}. Galaxy morphologies, nevertheless, mostly likely occupy a continuous space and discovery tools based on similarity\footnote{\url{https://mwalmsley-decals-similarity-similarity-papkyg.streamlit.app/}} might be more informative \citep[e.g.,][]{Walmsley2022MNRAS.513.1581W}.

Often the model hyperparameter choices are driven by prior knowledge and intuitive expectations based on highly biased and incomplete examples drawn from past datasets. We recommend to trust the data over models when the goal is discovery of new phenomena. \citet{Allen2024AnRSA..1140120A-interpretable-ML} in their review, discuss categories of discoveries, model intrepretability and validation in ML.

\section{Summary}

This review summarised the usage of Unsupervised Learning in Astronomy applications. Following a learning workflow, we summarised the most popular methods used in astronomy highlighting only some of the many thousands published works. We observed a few patterns: some works set out to use machine-learning to solve a well defined and specific problem, other works use a suite of algorithms on a common training dataset, others use some machine learning as part of larger analysis framework. Recently, a number of works combine machine-learning algorithms in very complex pipelines, some times with unclear advantages. The field is moving towards complex architectures, which aim to link domains in order to extract knowledge by leveraging past efforts.

However, no method is infallible. The choice and training of a machine-learning algorithm combined with implicit and explicit prior assumptions, introduced for example in the construction of the training set, harbour biases that might be difficult to diagnose. It is important that the scientific question is well defined before settling on any methods. Even though experimenting with new ML algorithms is useful, and certainly fun, domain knowledge shall remain the ultimate tool in the astronomer's toolbox.

\section*{Acknowledgements}
The author would like to acknowledge many interesting discussions with colleagues that predate the writing of this review: Johannes Buchner, Kartheik Iyer, Kai Polsterer, Crispin Logan, Grant Stevens, Myank Singhal. The author is grateful for her participation in the Kavli Summer Program in Astrophysics 2019 hosted at University California Santa Cruz, and the 2023 Kavli Institute for Theoretical Physics (KITP) programme ''Building a Physical Understanding of Galaxy Evolution with Data-driven Astronomy''. This research has made heavy use of NASA’s Astrophysics Data System.
\appendix

\clearpage

\onecolumn
\section{Method overview}
\begin{longtable}{p{0.35\textwidth}p{0.1\textwidth}p{0.2\textwidth}r}
\caption{Dimensional reduction methods discussed in this work} \label{tab:ML_dim_red_method_papers} \\
\hline
\hline \multicolumn{1}{c}{\textbf{Method}} & \multicolumn{1}{c}{\textbf{Acronym}} & \multicolumn{1}{c}{\textbf{Reference}}  \\ \hline 
\endfirsthead

\multicolumn{2}{c}%
{{\bfseries \tablename\ \thetable{} -- continued from previous page}} \\
\hline \multicolumn{1}{c}{\textbf{Method}} &\multicolumn{1}{c}{\textbf{Acronym}} & \multicolumn{1}{c}{\textbf{Reference}} \\ \hline 
\endhead

\hline \multicolumn{3}{|r|}{{Continued on next page}} \\ \hline
\endfoot

\hline \hline
\endlastfoot
Singular Value Decomposition & SVD & \citet{Press2007,Ivezic2014sdmm.book.....I} \\ \hline
Principal Component Analysis & PCA & \citet{Hotelling_1936_PCA} \\ \hline
Kernel principal component analysis & Kernel PCA & \citet{Scholkopf_1998_kernel_PCA} \\ \hline  
Independent Component Analysis & ICA & \citet[][review]{HYVARINEN2000-independent-component-analysis-review} \\ \hline  
Non-Negative Matrix Factorization & NNMF & \citet{Lee1999-NNMF}\\ \hline
Self-Organising Map & SOM & \citet{Kohonen1982_som} \\ \hline
Deep Embedded Self-Organising Map & DESOM & \citet{Forest2019-DESOM} \\ \hline
Auto-Encoder & AE & \citet{kramer1991_ae, KRAMER1992_AE} \\ \hline
Vector quantized Variational Autoencoder & VQ-VAE & \citet{van-den-oord-2017-vq-vae} \\ \hline
Gaussian Mixture Models & GMM & \citet{Dempster_1977_gaussian_mixture} \\ \hline
Isometric feature mapping & Isomap & \citet{Tenenbaum2000-ISOMAP} \\ \hline
Local Linear Embedding & LLE & \citet{Roweis2000-locally-linear-embedding}  \\ \hline 
t-distributed Stochastic Neighbor Embedding & tSNE & \citet{Maaten2008_tSNE} \\ \hline
Uniform Manifold Approximation and Projection for Dimension Reduction & UMAP & \citet{McInnes2018_UMAP} \\ \hline
  Simple Framework for Contrastive Learning of Visual Representations & simCLR & \citet{Chen2020-simCLR} \\ \hline
   Bootstrap Your Own Latent & BYOL & \citet{Grill2020_BYOL} \\ \hline
\end{longtable}

\begin{longtable}{p{0.35\textwidth}p{0.1\textwidth}p{0.2\textwidth}r}
\caption{Clustering methods discussed in this work} \label{tab:ML_clustering_method_papers} \\
\hline
\hline \multicolumn{1}{c}{\textbf{Method}} & \multicolumn{1}{c}{\textbf{Acronym}} & \multicolumn{1}{c}{\textbf{Reference}} \\ \hline 
\endfirsthead

\multicolumn{2}{c}%
{{\bfseries \tablename\ \thetable{} -- continued from previous page}} \\
\hline \multicolumn{1}{c}{\textbf{Method}} &\multicolumn{1}{c}{\textbf{Acronym}} & \multicolumn{1}{c}{\textbf{Reference}}  \\ \hline 
\endhead

\hline \multicolumn{3}{|r|}{{Continued on next page}} \\ \hline
\endfoot

\hline \hline
\endlastfoot

  k-means & & \citet{macqueen1967_k-means,lloyd1982_k-means}  \\ \hline
  k-medoids & & \citet{Ng2002-k-medoids-CLARANS,PARK2009-k-medoids} \\ \hline
  Hierarchical Clustering & HC & \citet{hastie_2009_elements-statistical-learning_book} \\ \hline
  Agglomerative clustering & AL & \citet{Johnson_1967_agglo_clustering} \\ \hline
  Minimum Spanning Tree\footnote{Known as Friends-of-Friends algorithm in astronomy.} & MST & \citet[][historical review]{Graham-1985-minimum-spanning-tree-historical-review}\\ \hline
  Density-based Spatial Clustering of Applications with Noise & DBSCAN & \citet{ester1996_dbscan} \\ \hline
  Density-based Spatial Clustering of Applications with Noise & HDBSCAN & \citet{Campello_2013_HDBSCAN, McInnes2017_HDBSCAN} \\ \hline
  Unsupervised Photometric Membership Assignment in Stellar Clusters & UPMASK & \citet{Krone-Martins2014-UPMASK} \\ \hline
 Isolation Forest & IF & \citet{Liu2008_Isolation_Forest} \\ \hline
  Fuzzy c-means & FCM & \citet{BEZDEK1984-fuzzy_c-means} \\ \hline
  Unsupervised fuzzy clustering & & \citet{gath-geva-1989-fuzzy-clustering}  \\ \hline
  Bootstrap Your Own Latent & BYOL & \citet{Grill2020_BYOL}\\ \hline
\end{longtable}
\clearpage
\twocolumn

\bibliographystyle{elsarticle-harv} 
\bibliography{UL-methods,SF-lib01,SF-lib02,SF-lib03,SF-lib04,SF-lib05,SF-lib06,SF-lib07,SF-mypapers}

\end{document}